\documentstyle[12pt,aasms4]{article}

\lefthead{Sembach  et al.}
\righthead{Highly Ionized HVCs}

\begin{document}

\newcommand{\chvcs}{{\ion{C}{4}}-HVCs}
\newcommand{\degr}{$^{\circ}$}
\newcommand{\et}{{\it et al.~}}
\newcommand{\hhvcs}{{\ion{H}{1}}-HVCs}
\newcommand{\kms}{\,km\,s$^{-1}$}     
\newcommand{\lam}{$\lambda$}
\newcommand{\lya}{Ly$\alpha$\ }
\newcommand{\subsun}{\mbox{$_{\odot}$}}
\newcommand{\twid}{\,$\sim$\,}
\newcommand{\vlsr}{V$_{\sc LSR}$}

\title{Highly Ionized High Velocity Clouds: Intergalactic Gas in the 
Local Group or Distant Gas in the Galactic Halo?\footnotemark}

\footnotetext{\noindent Based on observations obtained with the 
NASA/ESA Hubble 
Space Telescope, obtained at the Space Telescope Science Institute, which
is operated by the Association of Universities for Research in Astronomy, Inc.
under NASA contract NAS5--26555.}

\author{Kenneth R. Sembach}
\affil{Department of Physics \& Astronomy, The Johns Hopkins University,
Baltimore, MD  21218 \nl
{\it sembach@sundoggie.pha.jhu.edu}}

\author{Blair D. Savage}
\affil{Department Astronomy, The University of Wisconsin, Madison, WI 53706 \nl
{\it savage@astro.wisc.edu}}

\author{Limin Lu\footnotemark}
\affil{Department Astronomy, 105-24, California Institute of Technology 91125 \nl
{\it ll@troyte.caltech.edu}}
\footnotetext{Hubble Fellow}

\author{Edward M. Murphy}
\affil{Department of Physics \& Astronomy, The Johns Hopkins University,
Baltimore, MD  21218 \nl
{\it emurphy@pha.jhu.edu}}

\begin{abstract}
In the course of our studies of the gaseous halo surrounding the Milky Way, 
we have recently identified several high velocity (\vlsr\,$<$\,-100\kms) 
clouds in the directions of Mrk~509 and PKS~2155-304 that have unusual 
ionization properties.  The clouds exhibit strong \ion{C}{4} absorption with
little or no detectable low ion (\ion{C}{2}, \ion{Si}{2}) absorption or 
\ion{H}{1} 21\,cm emission down to very sensitive levels.  As the closest 
known analog to the outer diffuse halos of damped \lya absorbers and the 
low \ion{H}{1} column density metal line absorption systems seen in the 
spectra of high redshift quasars, these ``\chvcs'' present unique 
opportunities for relating the conditions within the Milky Way halo and 
nearby intergalactic gas to the properties of
galactic halos at higher redshift.

In this paper we present new Goddard High Resolution Spectrograph 
intermediate-resolution measurements of the absorption lines within these 
\chvcs\ and study the ionization properties of the gas in detail.  The present 
data represent the most complete set of measurements available for studying 
the ionization conditions within high velocity clouds.  The \chvcs\ have 
ionization properties consistent with photoionization by extragalactic 
background radiation, though some contribution by collisional ionization 
within a hot plasma cannot be ruled out.  The clouds are probably low density 
(n$_H$\twid10$^{-4}$~cm$^{-3}$), large (greater than several kiloparsecs), and 
mostly ionized (n$_{HI}$/n$_{H}$\twid10$^{-3}$) regions located well beyond 
the neutral gas layer 
of the Galaxy.   The presence of weak \hhvcs\ detected through their 21\,cm 
emission near both sight lines indicates that the \chvcs\ trace the 
extended, ionized, low density regions of the \hhvcs.  Several lines of 
evidence, including very low thermal pressures 
(P/k\twid2~cm$^{-3}$~K), favor a 
location for the \chvcs\ in the Local Group or very distant Galactic halo. 
Group.  
If the clouds are intergalactic in nature, their metallicities could be 
[Z/H]\twid-1 or lower, but higher metallicities [Z/H] $>$ -1 are favored 
if the 
clouds are located in the distant Galactic halo since the cloud sizes scale 
inversely with metallicity.  We provide a summary of the HVCs detected in 
absorption at intermediate resolution with the GHRS and the IUE satellite and 
find that \chvcs\ are detected along 3 of 10 extragalactic sight lines down 
to a level of log\,N(\ion{C}{4})\,$\approx$\,13.3 (3$\sigma$).

\end{abstract}


\section {Introduction}
Understanding the properties of high velocity clouds (HVCs) may prove to be an 
important step in revealing the processes that distribute and ionize gases 
in the halos of the Milky Way and other galaxies.  There have been many ideas 
proposed to explain the distribution of HVCs on the sky and their kinematics, 
including supernova-driven expulsion of hot gas into the Galactic halo
wherein the gas cools and returns to the disk in a 
global circulation pattern (i.e., a ``Galactic fountain'' -­ Shapiro \& 
Field 1976; Bregman 1980; Norman \& Ikeuchi 1989; Houck \& Bregman 1990), 
ram pressure or tidal stripping of material from the Magellanic Clouds 
(Moore \& Davis 1994; Lin, Jones, \& Klemola 1995), gas falling into the 
Milky Way (Oort 1970; Mirabel \& Morras 1984), and  
intergalactic clouds accreting onto the Local Group 
(Blitz \et 1998).  Available 
data suggest that each of these ideas may be appropriate for some HVCs, but 
that no single process can adequately describe the character of the entire 
ensemble of known HVCs (c.f., Wakker \& van Woerden 1997).

Many of the same processes that distribute high velocity gases within the 
Galaxy may also affect the underlying gases of the disk and halo.  For 
example, mass and energy are deposited into the interstellar medium (ISM) 
by stellar winds and supernovae, which evacuate large regions of the ISM by 
sweeping up and accelerating interstellar gas (c.f., Spitzer 1990; McKee 1993).
Supernovae therefore play a significant role in determining the distribution 
and ionization of the highly ionized gas in the low halo of the Milky Way 
(Sembach, Savage, \& Tripp 1997; Savage, Sembach, \& Lu 1997).  At larger 
distances from the Galactic plane, other processes impacting the properties 
of HVCs (e.g., tidal effects, interstellar-intergalactic medium interactions, 
ionization by extragalactic background radiation) may determine the larger 
scale distributions and physical properties of the gaseous Galactic halo.

Studying gas in the distant halo is observationally challenging since the 
spectral signature of the gas can be masked by the absorption due to 
foreground gas in the Galactic disk and low halo.  Therefore, studies of 
HVCs are particularly worthwhile since there is mounting evidence that 
many HVCs are located within the Galactic halo at large distances from the 
Galactic plane (c.f., Wakker \& van Woerden 1997).  By definition, HVCs have 
observed line-of-sight velocities $|$\vlsr$|$\,$>$\,100\kms, and though this 
velocity cutoff is somewhat arbitrary, it provides a convenient and 
historical context in which to classify clouds having velocities 
significantly exceeding those expected from typical random cloud motions 
and differential Galactic rotation (Spitzer 1978).

Using the Goddard High Resolution Spectrograph (GHRS) aboard the Hubble Space 
Telescope (HST), we recently discovered a pair of unusual high velocity 
clouds in the direction of Mrk~509 ($l$\,=\,36.0\degr, $b$\,=\,-29.9\degr).  
These HVCs were detectable only through their \ion{C}{4} absorption 
signature against the continuum of the Seyfert galaxy.  They showed no 
detectable low ionization (e.g., \ion{Si}{2}) absorption or \ion{H}{1} 
21\,cm emission at corresponding velocities (Sembach \et 1995b -­ hereafter, 
Paper I), as is usually seen for gases in the disk or low halo of the Milky 
Way (Sembach \& Savage 1992; Savage \et 1997).  We called these clouds 
\chvcs\ to distinguish them from the \hhvcs\ commonly detected in 21\,cm 
emission.  Mapping of the \ion{H}{1} emission near the Mrk~509 sight line 
revealed \hhvcs\ with similar velocities within $\sim$2\degr\ of the sight 
line, which led us to suggest that the \chvcs\ may be the ionized, extended, 
low density regions surrounding the \hhvcs.  

The unusual ionization properties of the clouds may provide insight into 
the properties of the distant Galactic halo, the processes that affect gas at 
large Galactocentric radii, and the origin(s) of HVCs.  Therefore, 
we initiated 
a two-part observing program to better understand the properties of the 
\chvcs.  We obtained additional intermediate-resolution GHRS measurements of 
several ions along the Mrk~509 sight line specifically to study the ionization 
conditions in the HVCs in detail.  We also identified a pair of \chvcs\ toward 
PKS~2155-304 ($l$\,=\,17.7\degr, $b$\,=\,-52.2\degr) that have ionization 
properties resembling those of the \chvcs\ toward Mrk~509.. 

We report on these new results in this paper, which is organized as follows:  
In  \S2 we describe the new observations and data reduction procedures.  
Section 3 contains the basic interstellar measurements and derived quantities 
such as ionic column densities.  Section 4 is an overview of the high 
velocity gas in the directions of Mrk~509 and PKS~2155-304.  In \S5 we 
model the ionization conditions in the \chvcs, and in \S6 we comment on the 
implications of the ionization results for the thermal pressures of the 
clouds and their distances.  In \S7 we discuss the relevance of this work for 
understanding damped \lya systems (galaxies) at high redshift.  
In \S8 we present a summary of the intermediate and high velocity 
gas detections made with the GHRS toward extragalactic objects.  Section 9 
contains our conclusions.

\section {Observations and Reductions}
\subsection{GHRS Ultraviolet Absorption Line Data}
We obtained GHRS intermediate resolution (G160M) observations of PKS~2155-304 
and Mrk~509 in October of 1995 and 1996, respectively, as part of HST guest 
observer programs GO-5889 and GO-6412.  A summary of the existing and new 
observations used in this study is contained in Table~1, where we list the 
HST identifications, wavelengths covered, dates of observation, detector 
substepping patterns, on-spectrum integration times, velocity shifts 
required to put the spectra into a common heliocentric velocity rest frame, 
and the ISM species detected.  The 1257--1293\AA\ spectra for the PKS~2155-304 
sight line were kindly provided by Dr. John Stocke (HST program GO-6593).  
We used the large (2.0\arcsec$\times$2.0\arcsec) science aperture (LSA) for 
all observations to reduce slit losses.  Unless otherwise indicated, we used 
standard carrousel rotation (FP-SPLIT = 4) and spectrum deflection 
(comb-addition = 4) procedures to reduce fixed pattern noise structure in each 
spectrum caused by irregularities in the detector window and photocathode 
response.  We used detector substepping patterns (Step-Patt = 4 or 5) to 
provide full diode array observations of the off-source backgrounds, which 
we averaged and subtracted from the on-spectrum data.  Substepping pattern 4 
provides two substeps per resolution element, which is one diode.  We rebinned
observations having a sampling interval of four substeps per diode 
(Step-Patt = 5) to two substeps per diode to improve the signal-to-noise ratio 
while still satisfying the Nyquist requirement of two independent samples 
across the 1-diode resolution element.

The post-COSTAR\footnotemark~LSA observations listed in Table~1 have 
instrumental profiles with velocity resolutions (FWHM) of approximately 
14\kms\ (1550\AA), 16\kms\ (1400\AA), and 18\kms\ (1250\AA).  The profiles 
have a Gaussian core containing $\approx$70\% of the light and a broad wing 
containing $\approx$30\% of the light (see Figure~4 of Robinson \et 1998).  
The 
1222--1259\AA\ observations of PKS~2155-304 and 1231--1269\AA\ observations 
of Mrk~509 were obtained in 1993 prior to the installation of COSTAR and have 
a narrow (FWHM\,$\approx$\,20\kms ) core containing $\approx$40\% of the 
spread function area and broad ($\approx\pm$70\kms ) wings containing 
approximately 60\% of the area. For more information about the performance of 
the GHRS see Brandt \et (1994) and Heap \et (1995).  Technical information 
about the instrument can be found in Soderblom \et (1994).

\footnotetext{COSTAR is the Corrective Optics Space Telescope Axial 
Replacement used to correct for the spherical aberration in the HST 
primary mirror.}

\subsection{Velocity Determinations}
The standard GHRS reduction software produces spectra with absolute wavelength 
calibrations accurate to approximately $\pm$1 diode ($\approx$15\kms ).  
To obtain more 
accurate relative velocity scales among the different exposures, we determined 
relative velocity shifts for the individual Mrk~509 spectra by requiring that 
the low velocity portions of the low ionization absorption profiles agree in 
velocity when aligned.  The set of \ion{Si}{2} lines covered in the individual 
spectra (\lam\lam1190.42, 1193.29, 1260.42, 1304.37, 1526.71) allowed a 
consistent velocity scale to be determined for all of the Mrk~509 observations,
except the integration covering the \ion{Si}{4} region, for which there is no 
nearby \ion{Si}{2} line.  Since there was no apparent shift of the 
low velocity 
\ion{Si}{4} profiles compared to the \ion{C}{4} profiles (for which we have a 
simultaneous \ion{Si}{2} \lam1526 measurement), we adopted the default 
heliocentric velocity scale provided by the standard processing for this 
observation.  Two observations required 10\kms\ shifts with respect to the 
others, which is within the nominal accuracy of 
the standard processing.  It is particularly important to note that the 
velocity shift of -10\kms\  determined for the 1300--1337\AA\ integration is 
identical to the shift determined if the low velocity \ion{O}{1} \lam1302, 
\ion{C}{2} \lam1334, or \ion{C}{2}$^*$ \lam1335 absorptions in the same 
spectrum are used in the low ion comparison instead of \ion{Si}{2} 
\lam1304 (i.e., there is no velocity ``stretch'' over the 37\AA\ interval 
covered by the observation).  The velocities of the \ion{Si}{2} line cores in 
the adopted velocity frame agree well with the velocity of the \ion{H}{1}
21\,cm emission toward Mrk~509 (e.g., Savage \et 1997).  We estimate an 
accuracy of better than 5\kms\ in the adopted absolute velocities for 
the Mrk~509 data.
 
We used the \ion{S}{2} \lam\lam1250, 1253, 1259 lines to tie the
HST observations for PKS~2155-304 into the \ion{H}{1} 21\,cm emission 
velocity reference in this direction.  The small adjustments made to the 
velocities ($\approx$5\kms) are well within the uncertainties of the velocity 
scales provided by the standard processing techniques.  A comparison of the 
strong \ion{Si}{2} \lam1260 
line with the \ion{Si}{2} \lam1526 line shows that there are no 
gross velocity shifts in the longer wavelength data.  We estimate an accuracy 
of $\approx$5--10\kms\ in the adopted absolute velocities for the
PKS~2155-304 data.
 
We converted the velocities of all spectra into the Local Standard of
Rest (LSR)\footnotemark~by applying shifts $\delta$v = 8.9\kms\ for 
Mrk~509 and $\delta$v = 1.9\kms\ for PKS~2155-304, where \vlsr = 
V$_{helio}$ + $\delta$v.  All subsequent velocities in this paper are 
referenced to the LSR.

\footnotetext{We use the reduction to the LSR given by Mihalas \& Binney
(1982), which assumes that the Sun is moving in the direction $l$\,=\,53\degr,
$b$\,=\,25\degr\ at a speed of 16.5\kms.  The LSR conversions for the two
directions considered are within 0.5\kms\ of the corrections based on 
``standard solar motion'' frequently used in radio work.}
 
In Figure~1 we show the full diode array spectra for the three Mrk~509 
observations that have not appeared elsewhere (see Table~1 for references to 
similar plots of the other spectra used in this study).  The individual 
interstellar lines are identified above each spectrum, as are the high 
velocity cloud components, when they are sufficiently strong to be visible.  
Continuum normalized versions of all the interstellar lines identified in the 
Mrk~509 and PKS~2155-304 spectra listed in Table~1 are shown in Figures~2 
and 3. 

\subsection{\ion{H}{1} 21\,cm Emission Data for the PKS~2155-304 Sight Line}
To complement the HST data for the PKS~2155-304 sight line, we obtained 
\ion{H}{1} 21\,cm emission spectra for the sight line and 35 nearby positions 
using the NRAO 140-foot telescope.  The data have an angular resolution of 
21\arcmin\ and a velocity resolution of 4\kms\ after Hanning smoothing.  The 
locations of these pointings listed in Table~2 are shown in Figure 4.  
A similar table and map for the Mrk~509 sight line can be found in Paper~I.  
We used a second order polynomial to determine the baseline for each spectrum.
In Figure~5 we show the 21\,cm spectrum for PKS~2155-304 and several nearby 
positions where \ion{H}{1} is detected at high velocities.  Only data for 
$|$\vlsr$|\,>$\,50\kms\ are shown because the emission at lower velocities is 
not completely removed from the spectrum by the $\pm$1.7\degr\ position 
switching (on-off source) used to detect the weak, high velocity features.  
Further details about the observational techniques and data reduction are 
given by Murphy, Lockman, \& Savage (1995).

\section{Interstellar Measurements}
\subsection{Ultraviolet Absorption Lines Toward Mrk~509 and PKS~2155-304}
We present the basic interstellar measurements for the Mrk~509 and 
PKS~2155-304 high velocity clouds in Tables~3 and 4.  Continua for all lines 
were low order ($\le$3) polynomials fitted to featureless continuum regions 
within several hundred \kms\ of the absorption lines.  For both sight lines, 
we divided the HVC absorption into two velocity intervals and made equivalent 
width and apparent column density estimates based upon integrations within 
these intervals.  The apparent column density is the integral of the apparent 
column density profile defined by 

$N_a = \int N_a(v) dv = \int \frac{m_ec/ \pi e^2}{f\lambda(\AA)} \tau_a(v) dv 
= \frac{3.768\times10^{14}}{f\lambda(\AA)} \int ln \frac{1}{I(v)} dv$ 
[atoms cm$^{-2}$ (\kms)$^{-1}$]~~~(1)

\noindent where I(v) and $\tau_a$(v) are the normalized intensity and 
apparent optical depth at velocity v (see Savage \& Sembach 1991 for a 
discussion of the construction and interpretation of apparent optical depth 
profiles).  The equivalent width and apparent column density errors in 
Tables~3 and 4 allow for statistical noise in the data and continuum 
placement uncertainties (see Sembach \& Savage 1992).  Background 
uncertainties are minimal ($<$1\%) in these data since the scattered 
light properties of the GHRS first-order gratings are excellent (Ebbets 1992).
 
The \chvcs\ have the following general properties, as determined by fitting a 
pair of Doppler-broadened Gaussian absorption components to the \lam1548 and 
\lam1550 lines for each sight line.  The fits, which are overplotted on the 
\ion{C}{4} \lam1548 lines shown in Figures~2 and 3, are characterized by a 
central velocity ($<$\vlsr$>$), a Doppler width (b) containing both thermal 
and turbulent contributions, and a column density (N).  The fitted profiles 
were degraded to the GHRS resolution by convolution with a single Gaussian 
instrumental profile having FWHM = 14\kms.

Mrk~509:~~~~~~~~~$<$\vlsr$>\,\approx-283\pm3$\kms~~~~b\,$\approx$\,25$\pm$2\kms~~~~log\,N(\ion{C}{4})\,$\approx$\,14.10

~~~~~~~~~~~~~~~~~~~~~$<$\vlsr$>\,\approx-228\pm6$\kms~~~~b\,$\approx$\,19$\pm$6\kms~~~~log\,N(\ion{C}{4})\,$\approx$\,13.56
\bigskip

PKS~2155-304: $<$\vlsr$>\,\approx-256\pm7$\kms~~~~b\,$\approx$\,33$\pm$5\kms~~~~log\,N(\ion{C}{4})\,$\approx$\,13.50

~~~~~~~~~~~~~~~~~~~~~$<$\vlsr$>\,\approx-141\pm9$\kms~~~~b\,$\approx$\,44$\pm$8\kms~~~~log\,N(\ion{C}{4})\,$\approx$\,13.50

\noindent The column densities derived from the fits are consistent with 
those obtained by integrating $N_a(v)$ for the weaker \ion{C}{4} 
\lam1550 line for PKS~2155-304.  However, the values of $N_a$ for the 
\lam1548 and \lam1550 lines for Mrk~509 derived through Eq. (1) indicate that 
some unresolved saturation is probably present in these profiles.  The 
2-component Gaussian fits to the lines do not provide much guidance for 
applying a saturation correction since the inferred b-values are larger than 
the instrumental function width (14\kms), and the S/N of the data do 
not warrant more complex fits.  We note, however, that a Doppler-broadened 
curve of growth applied to the equivalent widths listed in Table~3 yields 
column densities for the two components that are about 0.2 dex higher than 
the fit values or those inferred solely from the weak line results.  Given 
that this saturation correction is (potentially) large and uncertain, we will 
adopt the weak line $N_a$ values as lower limits in our analysis below.
 
Measurement of the HVC absorption is straight-forward in all but one case; 
the \ion{Si}{2} \lam1260.422 HVC feature between -330 and -190\kms\ toward 
PKS~2155-304 is blended with low velocity \ion{S}{2} \lam1259.519 absorption.  
A limit on the amount of \ion{Si}{2} HVC absorption within 
the \ion{S}{2} profile can be obtained by comparing the low velocity 
\ion{S}{2} line strength with the two ``clean'' \ion{S}{2} lines at 
1250 and 1253\AA.  A single-component Doppler-broadened curve of growth 
applied to these equivalent widths, W$_\lambda$(1250) = 90$\pm$13 m\AA\ and 
W$_\lambda$(1253) = 126$\pm$14 m\AA, 
yields  a column density of log\,N(\ion{S}{2})\,$\approx$\,15.3.  
The measured equivalent width of the 
\ion{S}{2} \lam1259 line is W$_\lambda$(1259) = 146$\pm$19 m\AA. Based on this 
measurement, the column density derived from the other two \ion{S}{2}
lines, and the equivalent width errors on all three \ion{S}{2} lines, we 
estimate that $<$50 m\AA\ (2$\sigma$) of high velocity \ion{Si}{2} is
present within the core of the \ion{S}{2} \lam1259 profile.  This 
corresponds to a \ion{Si}{2} HVC column density of 
log\,N(\ion{Si}{2})$_{HVC}$ $<$ 12.86 (2$\sigma$) for a Doppler parameter 
b $>$ 5\kms.
 
Table~5 contains a summary of the adopted HVC column densities.  These 
estimates are based on the individual apparent column density measurements 
listed in Tables~3 and~4.  Errors and limits are 2$\sigma$ estimates.  We 
adopted the conservative approach of quoting lower limits based on weak line 
results whenever the apparent column densities for multiple lines of a 
species indicated that a significant saturation correction might be 
warranted (e.g., \ion{C}{4} toward Mrk~509) or when only a single strong 
line of an ion was measured (e.g., \ion{Si}{3} at -283\kms\ toward Mrk~509).  
We adopted upper limits based on strong line results whenever a detection 
was marginal ($\le$2$\sigma$) (e.g., \ion{Si}{4} at -228\kms\
toward Mrk~509 and \ion{S}{2} at -256\kms\ toward PKS~2155-304). 

\subsection{H I 21\,cm Observations Toward Mrk~509 and PKS~2155-304}
We presented \ion{H}{1} 21\,cm emission data for the Mrk~509 sight line in
Figures~1 and 2 of Paper~I.  The observing methods, data reduction procedures, 
and measurement philosophy used in that work were similar to those presented 
below for the PKS~2155-304 sight line.

Using a deep, frequency switched \ion{H}{1} 21\,cm emission spectrum for 
PKS~2155-304, we estimate 4$\sigma$ upper limits of log\,N(\ion{H}{1}) $<$ 
17.73 for the two \chvcs\ along the sight line.  The baseline RMS of this 
spectrum is 2.3 mK.  We derived estimates of N(\ion{H}{1}) for this 
pointing and the nearby position-switched spectra from the baseline RMS and 
the following equation: 

\noindent N(\ion{H}{1}) = 1.942$\times10^{18}$ $\Delta$v T$_{max}$ 
[atoms cm$^{-2}$]~~(2)

\noindent where $\Delta$v is the FWHM of the best fit Gaussian profile in \kms\
(assumed to be 30\kms\ when no line is detected), T$_{max}$ 
is the peak brightness temperature of the line in K (assumed to be 4 times 
the baseline RMS when no line is detected).  In all cases, we assumed the 
\ion{H}{1} is optically thin.  Table~2 contains values of T$_{max}$, 
$\Delta$v (FWHM), $<$\vlsr$>$, and log\,N(\ion{H}{1}) for 35 positions 
within $\approx$2\degr\ of PKS~2155-304.
 
A schematic map of the positions with velocity information is shown in 
Figure~4 for the PKS~2155-304 sight line.  Filled symbols indicate high 
velocity gas detections; partially filled symbols indicate that the high 
velocity gas is broad and spread out in velocity rather than concentrated as 
a discrete narrow feature; crosses indicate null detections.  The numbers to 
the upper right of the symbols indicate the LSR velocities of the features 
detected.
\notetoeditor{Editor, please keep Figures 4 and 5 near each other,
preferably on the same page.}

\section{High Velocity Gas in the Directions of Mrk~509 and PKS~2155-304}
There are no known large HVC complexes in the general directions of Mrk~509 
or PKS~2155-304.  The closest ensemble of clouds to the Mrk~509 sight is 
the Galactic center negative (GCN) velocity group catalogued by Mirabel 
\& Morras (1984) and Wakker \& van~Woerden (1991).  The GCN clouds 
generally have peak brightness temperatures several times to 
an order of magnitude higher than those we measure for the \hhvcs\ near the 
Mrk~509 and PKS~2155-304 sight lines.  Some of the known GCN clouds 
have velocities close to those of \ion{H}{1} and \chvcs\ we detect.  For 
example, in the catalogue of Mirabel \& Morras (1984), GCN clouds with 
\vlsr\,=\,-266 to -187\kms\ are located within $\sim$5\degr\
of the Mrk~509 sight line, and a GCN cloud with \vlsr\,=\,-222\kms\
is located within $\sim$5\degr\ of the PKS~2155-304 sight line.
 
The origin of the GCN clouds is unknown, with suggestions ranging from 
low surface brightness galaxies (Cohen \& Mirabel 1978; Mirabel \& Cohen 1979)
to tidally disrupted gas in the Magellanic Stream in which the gas is flowing 
in toward the disk near the Galactic center (Mirabel 1981, 1982; Giovanelli 
1981; Wakker \& van~Woerden 1991).  The low column density \hhvcs\ 
that we detect near the Mrk~509 and PKS~2155-304 sight lines, which are 
separated by $\approx$26\degr\ on the sky, may be characteristic of 
lower density regions of a larger ensemble of HVCs associated with the GCN 
clouds.
 
The \hhvcs\ near both the Mrk~509 and PKS~2155-304 sight lines have weak 
emission, with maximum brightness temperatures of $\sim$50 mK and 
N(\ion{H}{1})\,$\le$\,2$\times$10$^{18}$ cm$^{-2}$ (two lower velocity clouds 
near -75\kms\ toward PKS~2155-304 have column densities several times higher).
The \hhvcs\ near the Mrk~509 sight line have central velocities and 
widths comparable to those of the higher velocity (-283\kms) \ion{C}{4}-HVC. 
There is no detectable \ion{H}{1} at the velocities of the lower velocity 
(-228\kms) \ion{C}{4}-HVC.  This situation is reversed for the PKS~2155-304 
sight line, where the nearby \hhvcs\ occur at velocities somewhat lower than 
those of the lower velocity (-140\kms) \ion{C}{4}-HVC.  Figure 5 shows that,
in many cases, the \ion{H}{1}-HVC detections have widths (FWHM) of 
$\approx$25--40\kms, similar to the widths of \chvcs\ 
($\approx$30--45\kms).  In several pointings, the lower velocity gas is spread 
out in velocity and no discrete narrow feature is observed (e.g.,
position B in Figure 5). 
 
The physical connection between the \chvcs\ and the \hhvcs\ is still 
debatable, though the relative proximity and velocity 
similarities of the two types of HVCs suggest that they are related.  
The simplest 
relationship would be one in which the \chvcs\ 
trace the extended, ionized, low density regions of \hhvcs\ at large 
distances from the Galactic plane as suggested in Paper~I.  
Such a situation has been modeled by Ferrara \& Field (1994), who
considered the 
ionization structure of clouds in a hot Galactic halo subjected to 
extragalactic background radiation.  Their model predicts a core-interface 
structure for various combinations of the total hydrogen column density and 
the linear size of the cloud (see their Figure 6).  In their model, the 
cloud core consists primarily of neutral gas (perhaps \hhvcs), while the 
interface region 
between the core and the surrounding hot gas (perhaps \chvcs)
is ionized primarily by 
photoionization by the extragalactic background. Thermal conduction 
and shocks could also be important factors in establishing 
the ionization conditions in such situations.
 	
Wolfire \et (1995) also considered the thermal and ionization structures of 
clouds embedded in a hot Galactic halo.  They found that a stable two-phase 
neutral gas structure consisting of cold, neutral cloud cores and warm, 
neutral cloud envelopes exist over a narrow range of pressures at a given 
height from the Galactic plane.  The cold cores exist only for z\,$\le$\,20 
kpc.  There is no convincing evidence in the ultraviolet absorption line 
data for cold cores in the \chvcs\ toward Mrk~509 or PKS~2155-304.  The 
strengths and widths of the \ion{H}{1}-HVC profiles 
are more suggestive of warm, low density gas than cold cores.  This is 
consistent with a location for the \chvcs\ at large distances from the 
Galactic plane.
 	
Benjamin \& Danly (1997) recently proposed a simple model of Galactic infall 
in which low density \hhvcs\  have velocities determined primarily by the 
drag forces between the clouds and the warm and hot gases of the Galactic 
halo.  In such a model, the terminal velocity of a HVC, v$_T$, is related to 
the observed velocity through the simple relation v$_T$ = v$_{obs} \csc|b|$.  
The velocities of the \chvcs\ toward Mrk~509 are probably too high to 
be explained by this simple model since the values of v$_T$ 
($\approx$450--550\kms) are comparable to the Galactic escape velocity of 
$\approx$500\kms (Binney \& Tremaine 1987).  The implied distances 
for the clouds in this simple model are so large that the assumptions used 
in the Benjamin and Danly analysis are invalid.  [A similar conclusion is 
obtained if one attributes some of the peculiar velocity to Galactic 
rotation since large velocities (\vlsr\,$\le$\,-80\kms) are reached only 
for very large distances (d\,$>$\,45 kpc)].  A similar conclusion holds for 
the PKS~2155-304 sight line, though the observed velocities are smaller.  The 
large negative velocities of the GCN clouds in this part of the sky led 
Mirabel \& Morras (1984) to conclude that the HVCs are probably infalling 
gas clouds located at large distances from the Galactic plane.  

\section{Ionization of the \chvcs}
With the HST, it is feasible to use QSOs and active galactic nuclei (AGNs) 
as background sources for HVC absorption line studies.  Our present study of 
the Mrk~509 \chvcs\ provides the most complete ionic information available for 
high velocity clouds outside the Galactic disk.
 
The absorption line signatures of the \chvcs\ toward Mrk~509 and PKS~2155-304 
are unlike those of most sight lines through the Galactic disk and low halo.  
Low velocity ($|$\vlsr$|$\,$<$\,100\kms) gas in the 
Galactic disk and low halo is generally characterized by strong low 
ionization absorption in addition to high ion absorption (see Figures~2 
and~3).  The strong \ion{Si}{2} \lam\lam1526, 1260 lines and the \ion{C}{2} 
\lam1334 line are usually much stronger than either
the \ion{Si}{4} \lam\lam1393, 1402 or \ion{C}{4} \lam\lam1548, 1550 
lines.  
N(\ion{C}{4})/N(\ion{C}{2})\,$>$\,1 in the \chvcs, compared to $\ll$1 in the 
Galactic disk and halo. Furthermore, the \chvcs\ have 
N(\ion{C}{4})/N(\ion{Si}{4})\,$>$\,5 and 
N(\ion{C}{4})/N(\ion{N}{5})\,$>$5--10, 
while gas within the disk or low halo typically has
N(\ion{C}{4})/N(\ion{Si}{4})\,$\approx$\,3.8$\pm$1.9 and 
N(\ion{C}{4})/N(\ion{N}{5})\,$\approx$4.0$\pm$2.4 (Sembach \et 1997).  
Strong departures from these average
high ion ratios are generally observed only for 
sight lines that pass through known radio continuum loops, and 
in one case  high values are found for an extended interarm 
region (Sembach 1994).  
The combination of unusual high ionization ratios coupled with the absence
of low ionization absorption is also atypical of high velocity gases in
the Vela supernova remnant (Jenkins, Wallerstein, \& Silk 1976, 1984) 
or star-forming regions like 
Carina (Walborn \et 1998).
To a large degree, the \chvcs\ 
resemble the low column density (log\,N(\ion{H}{1})\,$<$\,17) QSO metal line 
absorption systems, in which one often sees strong \ion{C}{4} (and sometimes 
\ion{Si}{4}) absorption but little low ionization absorption (Steidel 1990; 
Songaila \& Cowie 1996).

\begin{center}
\emph{Photoionization by Extragalactic Background Radiation}
\end{center}

Given that the \chvcs\ are likely to be at large distances from the Galactic 
plane, we explored their ionization properties assuming that the dominant 
ionization mechanism within the gas is photoionization by extragalactic 
radiation due to the integrated light of QSOs and AGNs.  
Using the photoionization code CLOUDY (v90.02; 
Ferland 1996), we modeled the HVCs as plane-parallel slabs of gas bathed from 
both sides in extragalactic background radiation with the QSO spectral energy 
distribution given by Madau (1992) and a mean intensity at the Lyman limit 
J$_{\nu}$(LL) = 1$\times$10$^{-23}$ erg cm$^{-2}$ s$^{-1}$ Hz$^{-1}$ sr$^{-1}$
(Donahue, Aldering, \& Stocke 1995; Haardt \& Madau 1996).  This corresponds 
to an ionizing photon density n$_\gamma$\,=\,4.3$\times$10$^{-7}$ ph cm$^{-3}$.
We ignored the effects of intervening clouds on the shape of the spectral 
energy distribution, and we assumed that the clouds had a uniform gas 
density.  We scaled the elemental abundances in the model to an overall 
metallicity, [Z/H], relative to the solar system abundances given by Anders 
\& Grevesse (1989), though the differential elemental abundance pattern in 
the model remained constant (i.e., we assumed no dust or local nucleosynthetic 
enrichment effects).\footnotemark
 
\footnotetext{We use the square-bracketed quantity [Z/H] to indicate the 
logarithm of metal abundances relative to solar abundances: 
[Z/H]\,=\,log\,(Z/H)--log\,(Z/H)$_\odot$.}

The ionization structure of a cloud in which the ionization is photon 
dominated is essentially determined by the ionization parameter 
[$\Gamma$ = (n$_\gamma$/n$_H$)\,$\propto$\,(J$_\nu$/n$_H$)], 
the ratio of ionizing 
photon density to total gas density (Bergeron \& Stasinska 1986).  For 
clouds that are optically thin [$\tau_{LL}$\,$<$\,1 or 
log\,N(\ion{H}{1})\,$<$\,17.2], a few scaling laws apply for fixed values 
of $\Gamma$.
\begin{center}
N(Z) $\propto$ (Z/H)$\times$N(\ion{H}{1})

N(\ion{H}{1}) $\propto$ n$_{HI}\times$D

n$_{HI} \propto$ n$_H\times$(n$_H$/J$_\nu$)

\end{center}

\noindent
where D is the cloud size.  Strictly speaking, the scaling laws are only 
approximate and depend on the metallicity of the gas, which affects the 
temperature of the gas and hence the ionization balance.  For [Z/H]\,$<$\,0, 
the above approximations are adequate for our purposes.  The scaling 
laws become almost exact at [Z/H]\,$<$\,-1 when metal cooling becomes less 
efficient.  At log\,N(\ion{H}{1})\,$>$\,17.2 ($\tau$(LL)\,$>$\,1), the clouds 
become self-shielded and the scaling laws break down.  
We note that the present epoch value of J$_\nu$(LL) is still uncertain.  
Since the ionization structure of the gas depends only on $\Gamma$, a change 
in J$_\nu$(LL) does not affect the relative column densities of the 
ions as long as the total gas density is changed accordingly.  For example, 
when J$_\nu$(LL) is lowered by  a factor of 2, the ionization structure 
(e.g., \ion{C}{1} : \ion{C}{2} : \ion{C}{3} : \ion{C}{4}) of the gas is 
unchanged if the gas density is also lowered by a factor of 2.  If the 
density is lowered, the 
absolute column densities will also be lowered unless the cloud 
size is increased accordingly.
 
We computed a set of models with various values of log\,N(\ion{H}{1}) and 
[Z/H] to determine the optimal set of parameters that satisfy the observed 
column densities and column density limits of the ions in the \chvcs\ toward 
Mrk~509 and PKS~2155-304.  In Figure~6 we plot the results of one such model 
for the Mrk~509 \chvcs\ at -283\kms\ (left panel) and -228\kms\ (right panel).
The symbols correspond to different ions as defined in the figure legend.  
The solid lines connecting the symbols indicate the ranges of log\,$\Gamma$
(or log\,n$_H$) that satisfy the individual observed ionic column density 
measurements and limits.  The heavy vertical line in each panel indicates the 
value of log\,$\Gamma$ (or log\,n$_H$) that satisfies all of the observed 
column density constraints simultaneously.  For the -283\kms\ cloud we find 
that a value of log\,$\Gamma$\,$\approx$\,-2.55 (log\,n$_H$\,$\approx$\,-3.82) 
for a model with log\,N(\ion{H}{1}) = 16.3 and [Z/H] = -0.5 satisfies the 
detections and upper limits over a very narrow range in $\Gamma$.  The main 
observational constraints are provided by \ion{C}{2}, \ion{C}{4}, 
\ion{Si}{4}, and \ion{N}{5} (see Figure~6).  For the -228\kms\ cloud we find 
that a slightly larger value of log\,$\Gamma$\,$\approx$\,-1.90 
(log\,n$_H$\,$\approx$\,-4.48) for a model with log~N(\ion{H}{1}) = 14.7 
and [Z/H] = -0.5 marginally satisfies all of the observable constraints.  
For this cloud, the \ion{Si}{3} column density provides a significant 
additional constraint.  In both cases, the metallicities can be varied 
provided that the relative ionic column densities are maintained by 
adjusting N(\ion{H}{1}) according to the approximate scaling laws listed 
above.\footnotemark

\footnotetext{In other words, a model with log\,N(\ion{H}{1})\,=\,16.8 and 
[Z/H]\,=\,-1 yields approximately the same metal line column 
density results as one with log\,N(\ion{H}{1})\,=\,16.3 and [Z/H]\,=\,-0.5.}

The cloud sizes in the standard models discussed above with 
[Z/H]\,$\approx$\,-0.5 range from roughly 4 to 30 kiloparsecs.  Lower 
metallicity models imply larger cloud sizes and somewhat larger temperatures 
than the solar metallicity models.  We can reasonably eliminate metallicities
[Z/H]\,$<$\,-2.0 since the implied cloud size for the -283\kms\ C IV-HVC
toward Mrk~509 would be $\ge$500 kiloparsecs.  Within the context of the
photoionization models, [Z/H]\,$>$\,-1 is favored if the clouds are
associated with the Milky Way (i.e., in the distant
Galactic halo).  An intergalactic location for the clouds would allow
for lower metallicities.
 
The amount of \ion{C}{4} in the HVCs toward PKS~2155-304 is about an order 
of magnitude lower than the amount observed in the higher velocity HVC 
toward Mrk~509.  The available data indicate that these \chvcs\ can be 
modeled successfully with parameters similar to those used for the Mrk~509 
\chvcs.  For example, both clouds can be modeled with the standard 
parameters used for the -283\kms\ HVC toward Mrk~509 provided that the 
ionization parameter is lower, log\,$\Gamma$\,$\approx$\,-3.00 
(log\,n$_H$\,$\approx$\,-3.38).  Figure~7 contains the model results for the 
PKS~2155-304 \chvcs.  Both clouds can also be adequately represented with 
lower column density models like those for the -228\kms\ cloud toward Mrk~509 
provided that log\,$\Gamma$\,$\approx$\,-2.20 (log\,n$_H$\,$\approx$\,-4.18).  
Observations of additional ions along the PKS~2155-304 sight line would help 
to characterize these clouds more completely. 
 
This study refines our earlier estimates of the ionization conditions within 
the \chvcs\ toward Mrk~509 (Paper~I).  The range of acceptable models
available with the earlier data is now much more tightly constrained, and 
our original conclusion that these clouds are probably large, low density, 
highly ionized regions remains intact.  Our new observations have allowed us 
to quantify these statements more fully and to determine that the two \chvcs\ 
along the sight line may have different ionization properties.  These 
differences are most likely due to differences in gas density rather than 
to viewing angle or partial shielding of one cloud by another since the 
optical depths of the clouds are small ($<$1).  This may also explain why some 
HVCs at large distances from the Galactic plane have no 
detectable \ion{C}{4} absorption.  If the gas density becomes too high, 
little \ion{C}{4} is created.  For example, HVC287.5+22.5+240 has no 
detectable \ion{C}{4} absorption, and interferometric \ion{H}{1} data show a 
concentrated cloud directly along the sight line, with 
N(\ion{H}{1})\,=\,8$\times$10$^{19}$ cm$^{-2}$ (Lu \et 1998; Wakker \et 1998).
 
In the above discussion we have not considered the effects of a non-solar 
relative abundance pattern that might result if dust is present within the 
HVCs.  Clear evidence for dust in a (presumably) distant HVC along the 
NGC~3783 sight line has recently been found by Lu \et (1998).  Of the 
elements considered in our study, Si would be most affected by gas-phase 
depletion onto dust grains.  The depletion of C, N, and S onto dust grains
is expected to be very small (less than a factor of 3) in these types of 
environments (Savage \& Sembach 1996).  An intrinsic non-solar relative 
abundance pattern could also affect our conclusions.  For example, the 
abundance pattern may be similar to that seen for metal-poor halo stars, 
in which Si is slightly overabundant relative to C and N by a factor 
of $\approx$2 (Wheeler, Sneden, and Truran 1989).  
For the Mrk~509 clouds, a super-solar metallicity of [Z/H]\,$\sim$\,0.7--1.0 
and a silicon gas-phase depletion relative to carbon [Si/C]\,$\sim$\,1.0 
allows values of log\,$\Gamma$ as low as $\approx$-2.95 
(log\,n$_H$\,$\approx$H\,-3.43) for the -283\kms\ cloud and 
$\approx$-2.68 (log\,n$_H$\,$\approx$H\,-3.70) for the -228\kms\ cloud.  
These values of $\Gamma$ are roughly 3--6 times higher than the values in 
the ``standard'' models discussed above but do not significantly change the 
main conclusions of this paper.  We present a summary of the properties of 
the \chvcs\ toward Mrk~509 derived from these photoionization models 
in Table~6.

\begin{center}
\emph{Photoionization by Starlight}
\end{center}

An additional source of ionizing photons that was not considered in the 
previous section is the integrated light emitted from hot stars within the 
Galaxy.  The subject of how photons from O-stars affect the ionization 
properties of the diffuse ionized gas (DIG) in the ISM has been studied 
recently by several groups, who find that a sufficient number of photons 
may leak out of the Galactic disk to ionize the ``Reynolds Layer'' 
(Miller \& Cox 1993; Domg\"{o}rgen \& Mathis 1994).  The fraction of photons 
that escape from the disk into the halo at a given Galactocentric 
radius depends upon the ionization, porosity, and dust distribution within 
the ISM, as well as the adopted number and distribution of hot stars in the 
Galaxy.  Dove \& Shull (1994) estimate an average number of Lyman continuum 
photons escaping through both the Galactic \ion{H}{1} layer and the top of 
the DIG layer to be $\Phi_{LyC}$(*)\,$\approx$\,1.5$\times$10$^6$ 
cm$^{-2}$ s$^{-1}$.   Measurements of  the H$\alpha$ emission 
from high velocity clouds in Complexes A, C, and M imply 
$\Phi_{LyC}$(*)\,$\approx$\,(1.3--4.2)$\times$10$^5$ cm$^{-2}$ s$^{-1}$ if 
the emission arises from photoionization of the clouds (Tufte, Reynolds, 
\& Haffner 1998).  All three complexes lie several kiloparsecs from the 
Galactic plane (see Wakker \& van~Woerden 1997).  These estimates are larger 
than the Lyman continuum flux from the extragalactic background in our 
photoionization model ($\Phi_{LyC}$(EB)\,$\approx$\,1.27$\times$10$^4$
cm$^{-2}$~s$^{-1}$) and suggests that internal Galactic sources contribute 
to the ionization properties of gas in the low Galactic halo.
 
A pure stellar ionizing spectrum of the type ionizing the DIG 
(T$_{eff}$\,$\approx$\,38,000 K; Domg\"{o}rgen \& Mathis 1994) 
does not produce 
enough \ion{C}{4} to match the \ion{C}{4} to \ion{Si}{4} ratios 
measured in the \chvcs.  The amount of \ion{Si}{4} relative to \ion{C}{4} 
would be about an order of magnitude higher than observed unless Si is more 
heavily depleted onto dust grains than C by a similar factor 
(Giroux \& Shull 1997 ­ see their Figure~4).  Also, the amount of doubly 
ionized ions, such as \ion{Si}{3}, relative to \ion{C}{4} in such a model 
is expected to be considerably larger than is observed (see Bregman 
\& Harrington 1983).  To further explore the effects of a harder 
radiation field than that of the DIG but softer than that of the pure QSO
 power law spectrum, we recalculated our photoionization models with a QSO 
spectrum filtered by intervening clouds (Haardt \& Madau 1996) combined with 
a T$_{eff}$\,=\,50,000 K stellar spectrum having 
J$_\nu$(LL)\,$\sim$\,10$^{-22}$ erg cm$^{-2}$ s$^{-1}$ Hz$^{-1}$ s$^{-1}$.  
The intensity at the Lyman limit due to the stellar portion of the 
spectrum is about 10 times higher than that of the extragalactic background
contribution.  The resulting ionizing spectrum has a small break at the Lyman 
limit and a large break at 4 Rydberg.  This hybrid model can reproduce the 
observed ionic column densities for the -283\kms\ cloud 
only if log\,$\Gamma$\,$\approx$\,-2.0, which is comparable to the value 
required for the pure QSO power law spectrum.  For the -228\kms\ cloud toward 
Mrk~509, it is necessary to incorporate a Si/C depletion of about 0.5 dex to 
make the model match the observed column densities.  We plot 
these results for the Mrk~509 \chvcs\ in Figure~8.  Acceptable hybrid models 
yield physical parameters for the clouds similar to 
those listed in Table~6. 

\begin{center}
\emph{Collisional Ionization}
\end{center}

An alternative production mechanism for the ionized species in the Mrk~509 
\chvcs\ is collisional ionization within a hot plasma.  
There are a host of predictions for the column densities expected under 
various collisional ionization situations, including turbulent mixing of cool 
and hot gases in shear flows (Shull \& Slavin 1994), magnetic 
conductive interfaces (Borkowski, Balbus, \& Fristrom 1990), supernova 
bubbles in late stages of evolution (Slavin \& Cox 1992, 1993), and 
radiatively cooling flows (Benjamin \& Shapiro 1998).  
A recent review of these hot gas theories and their relevance to production 
of highly ionized gases in the disk and halo has been given by Sembach \et 
(1997).  The most highly ionized gas lines observed in our study (\ion{Si}{4}, 
\ion{C}{4}, and \ion{N}{5}) cannot be used alone to completely rule 
out a collisional ionization origin for the \ion{C}{4} observed in the HVCs 
toward Mrk~509 and PKS~2155-304.  However, the absence of significant low 
ionization absorption associated with the high ion detections indicates that 
a collisional ionization origin for most of the ionized gas is 
unlikely.  Except under special circumstances, all of these collisional 
ionization models predict that some low ionization gas will be present along 
with the hot gas, and several of these by their 
very nature have spatially coincident regions of hot and cool gases.
 
In Table~7 we compare the ionic ratios observed for the two \chvcs\ 
toward Mrk~509 with typical values found for the low disk and halo, values 
predicted for a hot (T$_{max}$\,=\,10$^6$~K) radiatively cooling gas, and 
values predicted for the extragalactic background photoionization models 
discussed above.  Only a few of the observed constraints can be satisfied 
by the radiatively cooling gas calculations, whereas the photoionization 
models satisfy all of the listed ratios (as well as others that are not 
listed).  One caveat to bear in mind before eliminating collisional 
ionization for the \chvcs\ is that the gas may be very hot 
(T\,$>$\,5$\times$10$^5$ K) and not yet have had time to cool and recombine to 
produce the lower ion species.  The ionization fractions of the singly 
ionized species observed depend strongly upon the assumed stopping point in 
the cooling flow calculations (see Benjamin \& Shapiro 1998).
 
Future observations of the Mrk~509 and PKS~2155-304 sight lines would be 
valuable for further discrimination among the various ionization scenarios 
for the \chvcs. The Far Ultraviolet Spectroscopic Explorer (FUSE) will 
observe both sight lines within the next few years.  The FUSE bandpass 
(905--1195\AA) includes the \ion{O}{6} \lam\lam1031, 1037 doublet, which is 
the best diagnostic of hot ($>$10$^5$  K) collisionally ionized gas in the 
ultraviolet spectral region.\footnotemark~Most collisional ionization 
models predict 
N(\ion{O}{6})/N(\ion{C}{4}) $>$ 1 if the gas is hot (Sembach \et 1997), 
though some turbulent mixing layer models allow lower values if the 
post-mixed gas temperature in the cooling flow is not too high 
(Shull \& Slavin 1994).  If the absence of low 
ionization gas in the \chvcs\ is due to high temperature collisional 
ionization, the \ion{O}{6} absorption should be strong.  In the pure power 
law photoionization models considered here, the predicted \ion{O}{6} column 
density is comparable to that predicted for \ion{N}{5} and is about an order 
of magnitude lower than the \ion{C}{4} column density (see Figures~6 and~8).  
The predicted 
\ion{O}{6} column densities are much less in the hybrid QSO/stellar spectrum 
models considered.  Therefore, distinguishing between the observational
signatures of hot collisionally ionized gas and photoionized gas in the 
\chvcs\ should be feasible.

\footnotetext{The ionization potential of \ion{O}{6} is 113.9 eV, compared to 
47.9 eV for \ion{C}{4} and 77.5 eV for \ion{N}{5} (Moore 1970).}

\section{On the Origin and Distances of the \ion{C}{4}-HVCs}
The results of our investigation into the ionization properties of the \chvcs\ 
in the previous section lead to two important conclusions.  First, it appears 
that the principal ionization mechanism is photoionization by light with a 
relatively hard (QSO-like) spectrum, though some 
contributions due to collisional ionization or a relatively hot stellar-like 
spectrum are possible.  Second, the derived thermal pressures in the 
photoionization models are very small, P/k $\sim$ 2 cm$^{-3}$~K, compared to 
typical interstellar thermal pressures of $\approx$\,2000--3000 cm$^{-3}$~K 
in the Galactic disk.  Both conclusions lead us to believe that 
the \chvcs\ are associated with intergalactic gas in the Local 
Group or very distant gas in the Galactic halo.
 
A simple argument in favor of a large distance for the \chvcs\ can be made by 
considering the properties of the \hhvcs\ near the Mrk~509 sight line.  
The thermal pressure in the \hhvcs\ is given by P/k = n$_{HI}$T, where 
n$_{HI}$ is the total neutral gas density and T is the 
temperature of the clouds.  The pressure will be larger if the gas in the 
\hhvcs\ is partially 
ionized.  The pressure can be recast as an equation for distance given by 
d = size / $\theta$ = (N(\ion{H}{1})/n$_{HI}$) / $\theta$ = 
(P/k)$^{-1}$ (T/$\theta$) N(\ion{H}{1}), where $\theta$ is the angular extent 
of a spherical cloud in radians.  The widths of the \ion{H}{1} profiles for 
the \ion{H}{1}-HVCs indicate that T\,$<$\,2$\times$10$^4$~K, but a more 
realistic estimate of the temperature can be made by requiring that it be 
less than the temperature derived for the highly ionized gas regions, 
T\,$\le$\,10$^4$~K.  Adopting a temperature of T $\approx$ 10$^4$ K, a 
typical column density of N(\ion{H}{1}) = 2$\times$10$^{18}$ cm$^{-2}$, and 
conservative (large) angular size $\theta$\,$\approx$\,2\degr\ 
(see Paper~I), we find that d(kpc)\,$\approx$\,185\,(P/k)$^{-1}$.  
Assuming rough thermal pressure equilibrium between the \chvcs\ and the 
\hhvcs\ yields d\,$\approx$\,30--200 kpc.  Smaller values of $\theta$ increase 
this estimate, while smaller values of T decrease it.  If the pressure in 
the \ion{H}{1} emitting gas is larger than estimated above due to ionization 
effects, the distance will decrease.
 
The \ion{C}{4}-HVC properties are roughly consistent with a core-interface 
(\ion{H}{1} -- \ion{C}{4}) structure in the model proposed by Ferrara \& 
Field (1994).  Such clouds would have to be located at large distances from
the Galactic plane for the extragalactic radiation field to dominate their 
ionization characteristics.  The implied thermal pressures for the \chvcs\
are much smaller (2--3 orders of magnitude) than those predicted for 
multi-phase models of the Galactic halo at $|$z$|$ $<$ 10 kpc (e.g., 
Wolfire \et 1995).  
The \ion{C}{4}-HVC pressures can also be compared to the pressure expected
for an extended Galactic corona. An estimate of gas density at very 
large distances from the Galactic plane follows from the detection of H$\alpha$
emission from the Magellanic Stream (Weiner \& Williams 1996).  If the 
observed H$\alpha$ emission is produced by ram 
pressure heating as the Magellanic Stream moves through a hot Galactic 
corona, the detections provide information about the conditions in the 
corona at $\sim$50 kpc from the Galactic plane.  The 
Weiner \& Williams analysis implies that the coronal gas has 
T\,$\sim$\,1.7$\times$10$^6$~K with n$_H$\,$\sim$\,1$\times$10$^{-4}$
cm$^{-3}$ and P/k = 2.3n$_H$T $\sim$ 390 cm$^{-3}$~K, two orders of 
magnitude larger then the \ion{C}{4}-HVC thermal pressures.
(The factor of 2.3 arises from the assumption that the gas is fully ionized
and contains 10\% He by number.) 
 
Given the large inferred pressures and ionization signature for the 
\chvcs\, 
it is interesting to consider the possibility that the \chvcs\ are 
intergalactic clouds rather than 
entities within the Galactic halo.  Giovanelli (1981) argued that much of 
the GCN HVC complex 
(to which the \chvcs\ may be related) was probably not part of a population 
of intergalactic clouds based upon the cloud velocities and turbulent motions 
within the clouds; he assigned them to gas associated with tidal debris from 
the Magellanic Stream.  However, Blitz \et (1998) 
recently considered the intergalactic origin hypothesis in more detail and 
suggested that when a 
larger ensemble of HVCs is considered, some HVCs may 
be more properly 
considered members of the Local Group than clouds within the Milky Way.  They 
base this conclusion on the velocity centroid of the restricted HVC cloud 
ensemble considered; this centroid has the same kinematical radial velocity 
centroid as the Local Group.  However, not all HVCs fit 
well into the Blitz \et model, including the Magellanic Stream and Complexes 
A, C, and M.  HVC~287.5+22.5+240 fits the kinematical and positional 
aspects of the model, but it has a 
metallicity, [S/H] = -0.6, that is too high to be consistent with a 
primordial gas cloud (Lu \et 1998).  Still, many of the \chvcs\ properties 
suggest that an intergalactic location for some HVCs may be possible.  In 
this context, it is important to note that the \chvcs\ contain metals.
 
If the \chvcs\ are indeed intergalactic clouds, then the metallicity 
constraint of [Z/H] $>$ -1, which was derived in \S5 under the 
assumption that the clouds are located in the distant Galactic halo and 
have sizes no larger than a few tens of kiloparsecs, may be relaxed. 
Observations of absorption lines along closely paired lines 
of sight toward quasars show that low redshift (z $\sim$ 0.7) \lya clouds have 
extents of $\sim$500 kiloparsecs 
(Dinshaw \et 1995).  If the \chvcs\ are of similar nature and size, then 
metallicities of  [Z/H] $<$ -1 may be allowed.
  
The derived properties of the \chvcs\ are consistent with the properties 
found for the high redshift (z $\sim$ 3) \lya clouds; for a sample of about 
30 clouds, Songaila \& Cowie (1996) find 
that N(\ion{C}{4})/N(\ion{H}{1})\,$\approx$\,2$\times$10$^{-3}$ and 
log\,$\Gamma$\,$\approx$\,-1.9 for 
10$^{14}$\,$<$\,N(\ion{H}{1})\,$<$\,10$^{15}$ cm$^{-2}$, and 
N(\ion{C}{4})/N(\ion{H}{1})\,$\approx$\,3$\times$10$^{-3}$ 
and log\,$\Gamma$\,$\approx$\,-2.5 for 
10$^{15}$\,$<$\,N(\ion{H}{1})\,$<$\,10$^{17}$.  According to recent 
cosmological simulations of structure formation involving gas hydrodynamics 
(Petitjean, Mucket, \& Kates 1995; Zhang, Anninos, \& Norman 1995; Hernquist 
\et 1996; Miralda-Escud\'e \et 1996), \lya clouds at 2\,$<$\,z\,$<$\,4 with 
10$^{14}$\,$<$\,N(\ion{H}{1})\,$<$\,10$^{17}$ cm$^{-2}$ trace the diffuse gas 
in the filamentary and sheet-like structures that surround and connect 
galaxies.  The simulations indicate that the gas in the sheets and filaments 
gradually become denser concentrations that 
eventually form galaxies.  It is not unreasonable to think that some of this 
gas may have survived to the present day.  The \chvcs\ and some of the 
\hhvcs\ may be manifestations of such 
filamentary structures within the nearby universe (see Blitz \et 1998).

\section{Comparisons with High Redshift Galaxies}	
One of the main motivations for studying the distribution and physical 
properties of gas in the Galactic disk and halo is to understand the origin 
of intervening metal absorption line systems found in spectra of quasars at 
high redshifts.  In this regard, a comparison of the 
absorption properties of the Milky Way with those of the QSO damped \lya 
(DLA) systems may be particularly relevant since the latter are thought to 
arise from the high redshift counterparts of present-epoch galaxies and may 
possibly be the progenitors of spiral galaxies (Wolfe \et 1986; 
Prochaska \& Wolfe 1997b).  Such a comparison was carried out recently by 
Savage \et (1993a), who compared the equivalent width distributions of high 
and low ion absorption lines from the 
Milky Way toward extragalactic sight lines with those from DLA systems at a 
mean redshift $<z>$ = 2.4.  Although there are several factors that can 
bias the comparisons and influence the observed character of the absorption 
line system (see Savage 1988), the low ion absorption lines 
arising from the Milky Way disk and halo seem to have an equivalent width 
distribution similar to most DLA systems. However, Savage \et noted that the 
differences in the high ion absorption lines are more acute in that the 
\ion{C}{4} and \ion{Si}{4} absorption from the Galaxy is 
significantly weaker on average than that found in DLA systems (see their 
Figure~10). These comparisons reveal primarily the differences in the 
kinematics of the gas rather than differences in heavy element abundances 
since most of the absorption lines involved in the comparisons are 
saturated.  Indeed, we now know that DLA systems at $<z>$\,$\approx$\,2.4 
generally have metallicities a factor of 10--100 below solar (c.f., 
Pettini \et 1997; Lu \et 1996).
 
The increasing availability of high resolution observations of DLA absorption 
systems allows direct comparisons of the absorption line profiles of various 
species and the gas kinematics in the DLA systems and the Milky Way.  This
provides more information than simple equivalent width 
measurements.  In particular, echelle observations carried out 
with the Keck 10m telescopes (Wolfe \et 1994; Lu \et 1996; Prochaska 
\& Wolfe 1996, 1997a) show that \ion{C}{4} and \ion{Si}{4} absorption lines 
in DLA systems generally have profiles very different from those of low 
ionization species in both the component structure and the velocity extent of 
the absorption. This apparent dis-jointedness between high and low ionization 
characteristics was interpreted by those authors to mean that most of the high 
ion gas in DLA systems arises from regions that are physically distinct from 
the low ion gas, possibly from low 
density, ionized gaseous halos surrounding the neutral gas concentration. In
comparison, the high ion (\ion{C}{4} and \ion{Si}{4}) and low ion absorption 
lines in the Galactic disk and low halo tend to have 
roughly similar absorption profiles and velocity extents (Sembach \& Savage 
1992; Savage \et 1993b). The properties of the \chvcs\ may shed new light on 
the issue.  The \chvcs\, by definition, occur at velocities beyond those 
where disk and low halo gas absorption is usually seen and show relatively 
strong high ion absorption with little or no detectable low ion 
absorption. In this regard, the \chvcs\ toward Mrk~509 and PKS~2155-304 are 
qualitatively similar to those of DLA absorption line systems.  
Quantitatively, there are still large differences 
between the average Galactic absorption and DLA absorption, mainly because 
the covering fraction of Galactic sight lines with \chvcs\ is small (see  
\S8).  The difference could be due to (possibly) greater rates of gas 
infall and a higher extragalactic ultraviolet
radiation field intensity at high redshift.

\section{A Summary of High and Intermediate Velocity Gas Toward 
Extragalactic Objects}
Data from previous investigations of halo gas along extragalactic sight lines 
can be used to inventory the types of high velocity clouds encountered along 
complete paths through the Galactic disk and halo.  Savage \& Sembach (1996) 
and Wakker \& van~Woerden (1997) have compiled lists of \ion{H}{1}-HVC metal 
line detections.  Here, we provide a comparison of some of 
these results with detections / non-detections of highly ionized species.  We 
have summarized the HVC detections in Table~8, where we indicate whether the 
HVCs have been detected in \ion{H}{1} emission, low ion absorption, and 
\ion{C}{4} absorption.  Cases of HVC detections in the low 
ionization lines that have not had an integration to search for the presence 
of \ion{C}{4} are not considered here.  When possible, we have assigned the 
high velocity gas detections to identified HVCs or cloud complexes.  Details 
about the individual detections / non-detections can be found in the 
references cited in Table~8.

\subsection{High Velocity Clouds (HVCs)}
In the sample of objects that have high quality high ionization data 
obtained at intermediate resolutions with the GHRS or International 
Ultraviolet Explorer (IUE) satellite we find:
 
\noindent 1) Three sight lines with \chvcs, weak low ionization absorption,
and no detectable high velocity \ion{H}{1} 21\,cm emission 
(Mrk~509, PKS~2155-304, and H\,1821+643 at 
\vlsr\ = -213\kms).  For both the Mrk~509 and PKS~2155-304 sight lines there 
is high velocity 
\ion{H}{1} 21\,cm emission within 2š of the sight lines (see \S4). The 
vicinity of the H\,1821+643 sight 
line has not yet been sensitively searched for low column density \hhvcs. 
 
\noindent 2) One case of high velocity \ion{C}{4} absorption possibly 
associated with the \ion{H}{1} warp of the outer 
Galaxy (H~1821+643 at \vlsr = -120\kms).  \ion{H}{1} 21\,cm emission and 
strong low ion absorption are present over the velocity range 
covered by the \ion{C}{4} absorption.
 
\noindent 3) One \ion{H}{1}-HVC detected in both \ion{H}{1} 21\,cm 
emission and strong low ion absorption, with 
possible \ion{C}{4} absorption at the 2.7$\sigma$ level (Fairall 9).  In 
this case the HVC is a portion of the Magellanic Stream.
 
\noindent 4) Two cases of H I-HVCs seen in both \ion{H}{1} 21\,cm emission 
and low ionization absorption with no detectable \ion{C}{4} absorption 
to a 3$\sigma$ level of log\,N(\ion{C}{4})\,$\le$\,13.33 (Mrk~205 
and NGC~3783).
 
\noindent 5) Three cases where there are no obvious \ion{H}{1} or \chvcs\ 
(3C~273, NGC~3516, NGC~5548).  
These sight lines pass through very different regions of the Galaxy.  The
3C~273 sight line is a high latitude direction that passes through Loop~I
and near the edge of Loop~IV.  Intermediate 
velocity gas exists along the sight line, though some of the negative gas 
velocities are undoubtedly due to differential Galactic rotation 
(see Savage \et 1993b).  The NGC 3516 sight line is a moderate latitude 
direction that passes through Loop~III near HVC
Complex C.  The NGC~5548 sight line is a high latitude direction that passes
within 20\degr\ of the North Galactic Pole. 
 
\noindent 6) Numerous cases where high positive velocity 
absorption is seen in both low ionization and high ionization gas that is 
clearly associated with the LMC near +270\kms\ and the SMC near +150\kms.  The 
three best studied cases with the IUE satellite include HD~5980 in the SMC 
and HD~36402 and SN~1987A in the LMC.  GHRS results for \ion{C}{4} and 
\ion{Si}{2} have been published for two LMC stars by Bomans \et (1996).  
The ~270\kms\ velocity separation between the LMC and 
the Milky Way absorption permits a search for high velocity gas associated 
with the Milky Way in the velocity range from approximately +100 to +150\kms. 
de~Boer \et (1990) concluded that the absorption 
extending over this velocity range is not associated with the LMC.  While a 
location for this gas in the outer halo of the Milky 
Way is a distinct possibility, the kinematic complexity of Magellanic Cloud 
and Magellanic Stream gas along these directions makes it difficult to draw 
unambiguous conclusions.
 
\noindent 7) The inner Galaxy sight line to the star HD 156359, 
($l$ = 328.7\degr, $b$ = -14.5\degr, d = 11.1 kpc, and z = -2.8 kpc),
which has a 
HVC at +125\kms\ detected in \ion{C}{2}, \ion{Mg}{2}, 
\ion{Si}{2}, \ion{Si}{3}, and \ion{Fe}{2} 
with the IUE (Sembach , Savage, \& Massa 1991) and in
\ion{Si}{2} and \ion{N}{5} with the 
GHRS (Sembach, Savage, \& Lu 1995a).  In the direction of HD~156359, 
Galactic rotation should 
produce velocities ranging from 0 to -100\kms, while the HVC has a large 
positive velocity.  
In this particular case, the HVC was not detectable in the lines of 
\ion{C}{4} or \ion{Si}{4} with the IUE.  
The observations provide the constraints N(\ion{C}{4})/N(\ion{N}{5})\,$<$\,3.4 
and N(\ion{Si}{4})/N(\ion{N}{5})\,$<$\,0.9.
 
In all cases where the extragalactic sight line passes through a known 
\ion{H}{1}-HVC detected through \ion{H}{1} 21\,cm emission observations, 
ultraviolet low ionization absorption 
is present.  This result is not surprising since there has been no evidence 
for primordial (highly metal deficient) \ion{H}{1} high 
velocity gas clouds falling into the Milky Way. The 
observations noted here cover a wide range of Galactocentric distances and 
angular scales. 

\subsection{Intermediate Velocity Clouds (IVCs)}
Intermediate velocity (30\,$<$\,$|$\vlsr$|$\,$<$\,100\kms) gas 
exists along many of the 
extragalactic sight lines that have been observed with the GHRS.  Some of 
the more relevant cases for comparison with the high velocity gas include:
 
\noindent 1) Two \ion{H}{1}-IVCs with \ion{C}{4} absorption at similar 
velocities
(Mrk~205, NGC~3516).  In both cases, 
the \ion{C}{4} absorption may be due to a broader velocity distribution of 
high ion gas.  Component 
by component matching of the \ion{C}{4} absorption to the \ion{H}{1} 
emission is less clear than in the high velocity cases.
 
\noindent 2) One case of intermediate velocity \ion{C}{4} absorption at 
\vlsr\ = -70\kms\ that may be associated 
with the high-$|$z$|$ extension of the Perseus spiral arm (H~1821+643).  
\ion{H}{1} 21\,cm emission and low ion absorption are also seen at the 
velocities of the \ion{C}{4} absorption.
 
\noindent 3) Intermediate velocity ($\sim$\,60--90\kms) absorption in
\ion{C}{4}, \ion{Si}{4}, and lower ionization stages along most LMC 
sight lines observed with the IUE (Savage \& de~Boer 1981; Chu \et 1994; 
Bomans \et 1996). 

\subsection{8.3.  Sky Coverage}
Murphy \et (1995) found that approximately 37\% of sky is covered by \hhvcs\
down to a detection limit of log N(\ion{H}{1})\,$\approx$\,17.7 (5$\sigma$).  
Using the GHRS data for the extragalactic sight 
lines listed in Table~8, not including the LMC and SMC directions, we find 
\ion{C}{4} detections having log N(\ion{C}{4})\,$>$\,13.3 (3$\sigma$) with 
little or no associated low ionization absorption in 3 of 10 
cases.  Although the number of sight lines so far sampled is small, it 
appears that the possibility 
of detecting additional \chvcs\ through absorption line studies 
of this type along other sight 
lines is comparable to the chance of finding high velocity \ion{H}{1} 21\,cm 
emission along a sight line (4 of 10 cases).  Further detections of 
\ion{C}{4} and/or low ionization species at high velocities along 
other sight lines would allow better determinations of the relationships of 
the \chvcs\ and 
\hhvcs.  In principle, this information could be used to set a limit 
on the relative fraction of 
HVCs subject to ionization conditions like those toward Mrk~509 and 
PKS~2155-304.  Together with the known \chvcs, non-detections of \ion{C}{4} 
in the \hhvcs\ toward NGC~3783 and Mrk~205 show that ionization 
conditions and/or gas densities vary among HVCs. 

\section{SUMMARY}
We present new GHRS intermediate resolution (14--18\kms) observations of the 
absorption produced by high velocity gas in the directions of Mrk~509 and 
PKS~2155-304.  Both 
lines of sight have high velocity gas with unusual ionization conditions.  
The gas in these ``\chvcs'' exhibits strong \ion{C}{4} absorption with little 
or no low ion (\ion{C}{2}, \ion{Si}{2}) absorption or \ion{H}{1} 
21\,cm emission.  The \chvcs\ present unique opportunities to study the 
conditions within nearby intergalactic gas and the distant Milky Way halo.
The present study significantly extends our previous investigation into the 
properties of the \chvcs\ described in Paper~I.
 
The \chvcs\ have ionization properties that are very different from those of 
gases in the Galactic disk and low halo.  The ionic ratios are most consistent
with photoionization by the 
relatively hard extragalactic background radiation, though contributions by 
hot stellar sources or 
collisional ionization within a hot plasma cannot be ruled out.  The gas 
densities and cloud sizes 
are not well-determined due to uncertainties in the intensity of the
 extragalactic ultraviolet 
background and shape of the ionizing spectrum, but several general 
statements about the cloud properties can be made.  If the gas is 
photoionized by extragalactic background radiation or a 
combination of ultraviolet starlight and the extragalactic background, the 
clouds must be low 
density (n$_H$ $\sim$ 10$^{-4}$ cm$^{-3}$), large (greater than several 
kiloparsecs), and mostly ionized (n$_{HI}$/n$_H$ $\sim$ 10$^{-3}$) regions 
located well beyond the neutral gas layer of the Galaxy.  If the clouds are 
intergalactic in nature, their metallicities could be [Z/H] $\sim$ -1 or 
lower, but higher metallicities 
are favored if the clouds are located in the distant Galactic 
halo since the cloud sizes scale inversely with metallicity.  
 
Although we did not detect high velocity \ion{H}{1} emission at the positions 
of Mrk~509 or PKS~2155-304, weak \ion{H}{1} emissions were detected within 
2\degr\ of both sight lines at velocities similar to those of the \chvcs.  
The proximity of the \chvcs\ and \hhvcs\ in the sky 
indicates that the two types of high velocity gas may be related.  The 
simplest relationship would 
be one in which the \chvcs\ trace the extended, ionized low density regions 
of the \hhvcs. There is no clear association of the \ion{H}{1} or \chvcs\ 
detected in our searches with large HVC complexes on the sky, though the 
HVCs may belong to a group of high negative velocity \ion{H}{1} clouds known 
to exist in the direction of the Galactic center.
 
There are several lines of evidence suggesting that the \chvcs\ are at large 
distances 
from the Galactic plane and could possibly be intergalactic clouds in the 
Local Group: 1) The ionic line ratios in the \chvcs\ do not resemble those 
of clouds located within the Galactic 
disk or low ($|$z$|$ $<$ 5 kpc) Galactic halo.  Rather, the ionization 
properties of the clouds are more consistent with photoionization by 
extragalactic background radiation, which can only dominate 
the radiation field at large distances from the Galactic plane.  The 
ionization properties of the \chvcs\ are also similar to those of low 
N(\ion{H}{1}) metal line systems seen in QSO spectra, which 
are known to trace intergalactic gas.  2) The inferred thermal pressures 
in the \chvcs\ (P/k $\sim$ 2 cm$^{-3}$~K) are about three orders of magnitude 
lower than the thermal pressures found in the 
general interstellar medium in the Galactic disk or low Galactic halo.  3) 
A large distance (30--200 kpc) is implied if the \hhvcs\ near the \chvcs\
are in rough thermal pressure equilibrium with the \chvcs.  
4) There is an independent suggestion by Blitz \et (1998) that the \hhvcs\ 
in this region of the sky have a kinematical signature consistent with a 
location in the Local Group.

We have provided a summary of the HVCs detected in absorption at intermediate 
resolution with the GHRS and the IUE satellite.  Ignoring the sight lines to 
the LMC and SMC, we find that \chvcs\ are detected along 3 of 10 
extragalactic sight lines down to a level of
log\,N(\ion{C}{4})\,$\approx$\,13.3 (3$\sigma$).  The limited amount 
of data available indicates that the \chvcs\ and 
\hhvcs\ may have roughly similar sky covering factors.

Several measurements could be made to further reveal the nature of the 
\chvcs.  
Further ultraviolet observations of the PKS~2155-304 sight line with the 
HST would allow stronger 
constraints to be placed on the ionization conditions in the \chvcs\ along 
this sight line.  
Observations of \ion{O}{6} absorption with FUSE will determine whether 
collisional ionization is a 
viable mechanism for producing some of the \ion{C}{4} within these clouds.  
Identification and 
investigation of additional \chvcs\ would help constrain the typical 
kinematical properties 
and covering factors of the clouds, which could then be compared to higher 
redshift absorption 
systems.

Finally, deep H$\alpha$ imaging of the clouds would help to reveal the angular 
extents of the 
clouds and the spatial relationships between the \ion{H}{1} and \chvcs.  
By mapping out the H$\alpha$
emission in detail, it may be possible to set additional constraints on the 
distances and ionization properties of the clouds (Ferrara \& Field 1994; 
Bland-Hawthorn \et 1995; Bland-Hawthorn 1997; Tufte \et 1998).  For example, 
it may be possible to determine if 
collisional ionization due to ram pressure heating of the clouds as they 
move through a hot 
Galactic corona is an important contributor to the overall column densities 
of the ions observed 
in this study (particularly \ion{Si}{4}, \ion{C}{4},  and \ion{N}{5}).  We 
encourage those groups now performing 
sensitive H$\alpha$ emission measurements to observe the Mrk~509 and 
PKS~2155-304 regions of the sky.

\acknowledgments  
We appreciate helpful comments about HVCs and the manuscript provided 
by Robert Benjamin and Bart Wakker.  We thank Gary Ferland for providing a 
copy of the CLOUDY software and Francesco Haardt and Piero Madau for 
electronic versions of their ionizing spectra.  KRS acknowledges support from 
NASA Long Term Space Astrophysics grant NAG5-3485 and from grant 
GO-06412.01-95A from the Space Telescope Science Institute, which is operated 
by AURA under NASA contract NAS5-26555.  EMM acknowledges support 
from NASA through grant NAS5-32985.  BDS appreciates support from NASA 
through grant 
NAG5-1852.  LL recognizes support from Hubble Fellowship grant HF1062.01-94A.

\newpage
\noindent
Fig. 1. - Full diode array spectra of Mrk~509 for the 
1180--1216\AA, 1301--1337\AA, and 1383--1417\AA\ wavelength regions.  The 
interstellar lines in the spectra are identified, as are the 
high velocity cloud components associated with the \chvcs\ observed by 
Sembach \et (1995b).  An error spectrum is shown below each spectrum.  
The core of the \ion{O}{1} $\lambda$1302 line is 
affected by geocoronal \ion{O}{1} emission.  Similar illustrations of the 
other HST spectra for Mrk~509 
and PKS~2155-304 listed in Table~1 have been presented by Savage \et (1997).

\noindent
Fig. 2. - Continuum normalized interstellar lines observed toward Mrk~509.  
The data have velocity resolutions (FWHM) of 14--18\kms\ and typical continuum 
signal-to-noise ratios of $\approx$10 per resolution element.  The 
pre-COSTAR data between 1231 and 1269\AA\ have a narrow spread function core 
(FWHM $\approx$ 20\kms) and broad wings ($\approx \pm$70\kms).  The data in 
the 1230--1260\AA\ region has considerably higher S/N ($\approx$\,20--30) 
due to the elevation in flux caused by the \lya emission of the Seyfert 
galaxy. For some lines, additional absorption due to other species can be 
seen nearby; these features are identified in the figure.  For \ion{N}{1}, 
the zero velocity wavelengths of the three lines in the 1200\AA\ triplet
are indicated by vertical tick marks above the spectrum.  Note the presence 
of high velocity clouds (HVCs) at velocities between -340 and -170\kms\ in 
the lines of \ion{Si}{3}, \ion{Si}{4}, \ion{C}{4}, and possibly \ion{C}{2}.

\noindent
Fig. 3 -  Continuum normalized interstellar lines observed toward 
PKS 2155-304.  The data have velocity resolutions (FWHM) of 14--18\kms\ and 
continuum signal-to-noise 
ratios of $\approx$\,14--20 per resolution element.  The pre-COSTAR data 
shortward of 1259\AA\ have a narrow spread 
function core (FWHM $\approx$ 20\kms) and broad wings ($\approx \pm$70\kms).  
Note the high velocity clouds (HVCs) at -256 and -140\kms\ observed in the 
\ion{C}{4} lines.  The lower velocity HVC is 
also seen in the strong \ion{Si}{2} $\lambda$1260 line, but the 
higher velocity cloud is unobservable in the \ion{Si}{2} 
$\lambda$1260 line due to blending with low velocity \ion{S}{2} 
$\lambda$1259 absorption.

\noindent
Fig. 4 - Galactic coordinate map of the \ion{H}{1} 21\,cm emission 
pointings 
obtained for the general direction of PKS~2155-304 ($l$ = 17.7\degr, $b$ = 
-52.2\degr).  Filled symbols indicate high velocity \ion{H}{1}
detections; partially filled symbols indicate that the high velocity 
\ion{H}{1} is partially spread out in 
velocity rather than concentrated as a discrete narrow feature; crosses 
indicate directions with no 
detectable high velocity  \ion{H}{1} emission.  The numbers to the upper 
right of some symbols indicate 
the LSR velocities of the features detected.  The spectra for pointings 
A--F are shown in Figure~5.

\noindent
Fig. 5 - Position-switched \ion{H}{1} 21\,cm emission spectra for the 
PKS~2155-304 sight line and 
several neighboring sight lines.  The brightness temperature is indicated 
for each spectrum on the 
vertical axis.  Data points for velocities $|$\vlsr$| <$ 50\kms\ 
have been omitted since the $\pm$1.7\degr\ 
position switching does not completely remove the strong signal near zero 
velocity.  These 
NRAO 140-foot data have an angular resolution of 21\arcmin\ and a velocity 
resolution of 4\kms\ after 
Hanning smoothing.  The positions for pointings A--F are indicated in the 
map shown in Figure~4.  
A stray radiation corrected spectrum showing the emission at low velocities 
directly along the 
sight line can be found in Lockman \& Savage (1995).

\noindent
Fig. 6 - Photoionization model calculations for the ions 
observed toward the Mrk~509 high 
velocity clouds.  The symbol legend contains the ion-symbol identification, 
and the two numbers 
immediately below the HVC designation indicate respectively the neutral 
hydrogen column 
density log\,N(\ion{H}{1}) and the gas metallicity [Z/H].  The mean 
ionizing background intensity at the 
Lyman limit is assumed to be 1$\times$10$^{-23}$ erg cm$^{-2}$ s$^{-1}$ 
Hz$^{-1}$ sr$^{-1}$, though the relative column 
densities of the ions at a given value of $\Gamma$ do not depend on the value 
of J$_\nu$(LL) as long as the 
particle density is changed accordingly.  Solid lines indicate the portions 
of the ion curves that 
satisfy the observational constraints on the detections of \ion{C}{2}, 
\ion{C}{4}, and \ion{Si}{4}, and the upper 
limits on the remaining ions.  The heavy vertical line indicates the 
value of log\,$\Gamma$ (or log n$_H$) 
that allows a simultaneous fit to all of the observational constraints.  The
dashed line in each 
panel represents the predicted behavior of \ion{O}{6} 
(not observed in this study) for comparison with 
the other high ionization lines.

\noindent
Fig. 7 - Same as Figure~6, except for the PKS~2155-304 sight line.  Solid 
lines indicate the portions of the ion curves that satisfy the observational 
constraints for the \ion{C}{4} and \ion{Si}{2} detections and the upper 
limits for the other ions.

\noindent
Fig. 8 - Same as Figure~6, except for a QSO power law ionizing spectrum 
combined with a T$_{eff}$ = 50,000 K stellar spectrum.  In the right hand 
panel, the abundance of Si relative to C has been 
adjusted by -0.5 dex so that a suitable fit could be found.  The best fit 
values of $\Gamma$ inferred from these panels are similar to those shown in 
Figure~6.

\end{document}


\newcommand{\chvcs}{{\ion{C}{4}}-HVCs}
\newcommand{\degr}{$^{\circ}$}
\newcommand{\et}{{\it et al.~}}
\newcommand{\hhvcs}{{\ion{H}{1}}-HVCs}
\newcommand{\kms}{\,km\,s$^{-1}$}     
\newcommand{\lam}{$\lambda$}
\newcommand{\lya}{Ly$\alpha$\ }
\newcommand{\subsun}{\mbox{$_{\odot}$}}
\newcommand{\twid}{\,$\sim$\,}
\newcommand{\vlsr}{V$_{\sc LSR}$}

\setcounter{page}{46}

\begin{deluxetable}{lcccccccc}
\tablewidth{0pt}
\tablecaption{GHRS Observations of Mrk 509 and PKS 2155-304}
\tablehead{\colhead{Object}       & \colhead{Archive} 
         & \colhead{Wavelength}   & \colhead{Date}     
         & \colhead{Stepping}     & \colhead{Exp.} 
         & \colhead{Velocity}     & \colhead{ISM Species}     
         & \colhead{Original} \\
         & \colhead{ID\tablenotemark{a}}  & \colhead{Coverage} & 
         & \colhead{Pattern\tablenotemark{b}}& \colhead{Time\tablenotemark{c}} 
         & \colhead{Shift\tablenotemark{d}}  & \colhead{Detected} 
         & \colhead{Reference\tablenotemark{e}} \\
	 &                                   & \colhead{(\AA)}               
         &                                   &
         & \colhead{(sec)}                   & \colhead{(\kms)}              
         &                                   & } 
\startdata
Mrk~509  & Z1790208M  & 1231.8--1269.0  & \phn5/93 & 4 & 6144.0 &        $+10$  & \ion{N}{5}, \ion{S}{2}, \ion{Si}{2}           & 2,3 \nl
         & Z26O0108T  & 1521.2--1557.5  & \phn8/94 & 4 & 7168.0 &  \phs\phn$0$  & \ion{Si}{2}, \ion{C}{4}                       & 2,3 \nl
         & Z3IA0104T  & 1178.8--1216.1  &    10/96 & 5 & 2867.2 &  \phs\phn$0$  & \ion{Si}{2}, \ion{Si}{3}                      & 1 \nl
         & Z3IA0105T  & 1178.8--1216.1  &    10/96 & 5 & 2867.2 &  \phs\phn$0$  & \ion{Si}{2}, \ion{Si}{3}                      & 1 \nl
         & Z3IA0107T  & 1300.5--1337.5  &    10/96 & 5 & 3379.2 &        $-10$  & O I, \ion{Si}{2}, \ion{C}{2}, \ion{C}{2}$^*$  & 1 \nl
         & Z3IA0108T  & 1381.9--1418.6  &    10/96 & 5 & 4901.6 &  \phs\phn$0$  & \ion{Si}{4}                                   & 1 \nl \\
 
PKS~2155-304  & Z1AW0106T  & 1222.7--1258.8 & \phn5/93  & 4  & 1574.4  &  $-$\phn$5$  & \ion{S}{2},  \ion{N}{5}?  & 3   \nl
              & Z1AW0107T  & 1222.7--1258.8 & \phn5/93  & 4  & 1574.4  &  $-$\phn$5$  & \ion{S}{2},  \ion{N}{5}?  & 3   \nl
              & Z1AW0108M  & 1222.7--1258.8 & \phn5/93  & 4  & 1536.0  &  $-$\phn$5$  & \ion{S}{2},  \ion{N}{5}?  & 3   \nl
              & Z3E70504T  & 1257.6--1293.6 &    10/96  & 5  & 4080.0  &  $+$\phn$5$  & \ion{S}{2},  \ion{Si}{2}  & 1,4 \nl
              & Z3E70505T  & 1257.6--1293.6 &    10/96  & 5  & 3400.0  &  $+$\phn$5$  & \ion{S}{2},  \ion{Si}{2}  & 1,4 \nl
              & Z2WS0107P  & 1519.8--1556.1 &    10/95  & 4  & 4915.2  & \phs\phn$0$  & \ion{Si}{2}, \ion{C}{4}   & 3   \nl
              & Z2WS0108P  & 1519.8--1556.1 &    10/95  & 4  & 1024.0  & \phs\phn$0$  & \ion{Si}{2}, \ion{C}{4}   & 3   \nl
\tablenotetext{a}{All observations were obtained in the large science 
aperture using grating G160M.  Comb-addition and FP-SPLIT=4 were used in all 
cases, except for the PKS~2155-304 1222--1293\AA\ integrations.  Observations 
beginning with ``Z1'' are pre-COSTAR spectra, which have a narrow 
core containing $\approx$40\% of the spread function area and broad wings 
containing $\approx$60\% of the area.  Observations beginning with ``Z2'' 
or ``Z3'' are post-COSTAR spectra with spread functions having
$\approx$70\% of the area contained within a narrow core (see text).}

\tablenotetext{b}{Detector substepping pattern used.  STEP-PATT=4 indicates 
full-diode array spectra of length 1000 points with 2 samples per diode and 
$\approx$11\% overhead spent on measuring interorder background levels on 
regions of the detector adjacent to the spectrum.  STEP-PATT=5 indicates 
full-diode array spectra of length 2000 points with 4 samples per diode and 
$\approx$6\% overhead spent on measuring interorder background levels.}

\tablenotetext{c}{On-spectrum exposure time.  This time does not include 
time spent on background measurements.}

\tablenotetext{d}{The velocity shift required to put the spectra onto a common 
heliocentric velocity scale.  All shifts are within the uncertainty 
($\pm$1 diode) of the standard GHRS wavelength calibrations.}

\tablenotetext{e}{Original references in which full-diode array spectra have 
appeared: 1) This paper; 2) Paper~I; 3) Savage \et 1997; 4) Data kindly 
provided by Dr. John Stocke in advance of public release into the HST archive.}
\enddata
\end{deluxetable}

\clearpage
\begin{deluxetable}{cccccccc}
\tablewidth{0pt}
\tablecaption{NRAO 21\,cm Line Profile Results and H\,I Column Densities 
Near the PKS~2155-304 Sight Line\tablenotemark{a}}
\tablehead{\colhead{$l$} & \colhead{$b$} 
	& \colhead{North\tablenotemark{b}} & \colhead{East\tablenotemark{b}} 
	& \colhead{T$_{max}$\tablenotemark{c}} & \colhead{FWHM}
	& \colhead{$<$\vlsr$>$} & \colhead{N(\ion{H}{1})\tablenotemark{d}} \\
	  \colhead{(\degr)} & \colhead{(\degr)} 
	& \colhead{(\arcmin)} & \colhead{(\arcmin)}
	& \colhead{(mK)} & \colhead{(\kms)}
	& \colhead{(\kms)} & \colhead{(10$^{17}$ cm$^{-2}$)}}
\startdata
17.73 & $-$52.24 &   \phs\phn0 &  \phs\phn0  & $<$9.2 & \nodata & \nodata & $<$5.4 \nl
\nl
17.45 &	$-$50.77 &	\phs\phn0 &	\phn$-$89 &	$<$14.0 &	\nodata & 	\nodata &	$<$8.2 \nl	
17.59 &	$-$51.48 &	\phs\phn0 &	\phn$-$46 &	$<$13.6 &	\nodata &	\nodata &	$<$7.9 \nl	
17.66 &	$-$51.88 &	\phs\phn0 &	\phn$-$22 &	$<$12.8 &	\nodata &	\nodata &	$<$7.5 \nl	
17.80 &	$-$52.61 &	\phs\phn0 &	\phn+22 &	$<$11.6 &	\nodata &	\nodata &	$<$6.8 \nl	
17.86 &	$-$52.96 &	\phs\phn0 &	\phn+43 &	$<$12.4 &	\nodata &	\nodata &	$<$7.2 \nl	
17.93 &	$-$53.35 &	\phs\phn0 &	\phn+67 &	$<$12.0$^e$ &	\nodata &	\nodata &	$<$7.0 \nl		
17.99 &	$-$53.72 &	\phs\phn0 &	\phn+89 &	15.7$\pm$1.0 &	23.0$\pm$1.9 &	$-$113.9$\pm$0.7 &	7.0$\pm$0.7 \nl
18.06 &	$-$54.09 &	\phs\phn0 &	+111 &	9.0$\pm$1.1$^f$ &	54$\pm$13$^f$ &	$-$132.4$\pm$3.7 &	$<$9.4$^f$ \nl	
18.11 &	$-$54.43 &	\phs\phn0 &	+132 &	$<$12.1$^e$ &	\nodata &	\nodata &	$<$7.2 \nl	
18.18 &	$-$54.82 &	\phs\phn0 &	+155 &	$<$12.0$^e$ &	\nodata &	\nodata &	$<$7.0 \nl	
18.23 &	$-$55.14 &	\phs\phn0 &	+175 &	64.5$\pm$2.6 &	24.6$\pm$1.4 &	$-$77.5$\pm$0.6 &	31$\pm$2 \nl
18.29 &	$-$55.56 &	\phs\phn0 &	+200 &	147.2$\pm$2.0 &	33.8$\pm$0.6 &	$-$74.9$\pm$0.2 &	97$\pm$2 \nl
\nl
18.23 &	$-$50.71 &	+30       &	\phn$-$89 &	$<$16.0 &	\nodata &	\nodata &	$<$9.3 \nl
18.54 &	$-$52.19 &	+30       &	\phn\phs\phn0 &	$<$14.0 &	\nodata &	\nodata &	$<$8.2 \nl
18.83 &	$-$53.66 &	+30       &	\phn+89 &	15.1$\pm$1.0$^f$ &	20.3$\pm$1.8 &	$-$116.6$\pm$0.7 &	6.0$\pm$0.7$^f$\nl
\nl
19.02 &	$-$50.64 &	+60       &	\phn$-$90 &	$<$11.6 &	\nodata &	\nodata &	$<$6.8 \nl
19.35 &	$-$52.12 &	+60       &	\phn\phs\phn0 &	$<$12.0 &	\nodata &	\nodata &	$<$7.0 \nl
19.67 &	$-$53.60 &	+60       &	\phn+90 &	$<$12.8 &	\nodata &	\nodata &	$<$7.5 \nl
\nl
16.92 &	$-$50.07 &	$-$15     &	$-$133 &	$<$17.2 &	\nodata &	\nodata &	$<$10 \nl
17.19 &	$-$51.54 &	$-$15     &	\phn$-$44 &	$<$15.6 &	\nodata &	\nodata &	$<$9.1 \nl
17.33 &	$-$52.27 &	$-$15     &	\phn\phs\phn0 &	14.1$\pm$1.4$^f$ &	16.1$\pm$2.1$^f$ &	$-$132.9$\pm$1.0 &	$<$11$^f$ \nl	
17.45 &	$-$53.01 &	$-$15     &	\phn+44 &	$<$17.6 &	\nodata &	\nodata &	$<$10 \nl
17.57 &	$-$53.74 &	$-$15     &	\phn+89 &	$<$20.8 &	\nodata &	\nodata &	$<$12 \nl
\nl
16.66 &	$-$50.83 &	$-$30     &	\phn$-$88 &	$<$16.8 &	\nodata &	\nodata &	$<$9.8 \nl
16.92 &	$-$52.30 &	$-$30     &	\phn\phs\phn0 &	11.4$\pm$1.3$^f$ &	42.3$\pm$5.6$^f$ &	$-$110.2$\pm$2.4 &	$<$11$^f$ \nl	
17.15 &	$-$53.76 &	$-$30     &	\phn+88 &	16.4$\pm$1.8$^f$ &	46.5$\pm$6.2$^f$ &	$-$113.7$\pm$2.6 &	$<$15$^f$ \nl	
\nl
16.27 &	$-$50.86 &	$-$45     &	\phn$-$88 &	$<$16.0 &	\nodata &	 \nodata &	$<$9.3 \nl
16.51 &	$-$52.32 &	$-$45     &	\phn\phs\phn0 &	25.5$\pm$1.1 &	31.8$\pm$1.6 &	$-$117.9$\pm$0.7 &	16$\pm$1 \nl
16.73 &	$-$53.79 &	$-$45     &	\phn+88 &	18.6$\pm$0.8$^f$ &	87.2$\pm$5.2$^f$ &	$-$94.1$\pm$2.0 &	$<$31$^f$ \nl
\nl
15.88 &	$-$50.89 &	$-$60     &	\phn$-$88 &	$<$12.8 &	\nodata	& \nodata &	$<$7.5 \nl
16.10 &	$-$52.35 &	$-$60     &	\phn\phs\phn0 &	$<$12.0 &	\nodata	& \nodata &	$<$7.0 \nl
16.31 &	$-$53.81 &	$-$60     &	\phn+88 &	19.6$\pm$1.0 &	33.9$\pm$2.3 &	$-$88.3$\pm$0.9 &	13$\pm$1 \nl
\nl
15.09 &	$-$50.94 &	$-$90     &	\phn$-$87 &	$<$15.6 &	\nodata &	\nodata &	$<$9.1 \nl
15.29 &	$-$52.39 &	$-$90     &	\phn\phs\phn0 &	$<$14.8 &	\nodata &	\nodata &	$<$8.6 \nl
15.46 &	$-$53.84 &	$-$90     &	\phn+87 &	$<$16.4 &	\nodata &	\nodata &	$<$9.6 \nl

\enddata
\end{deluxetable}

\clearpage
\begin{center} TABLE~2 NOTES \end{center}

$^a$ A similar table for the Mrk 509 sight line can be found in Sembach \et
(1995b ­ see their Table~2).

$^b$ Beam position relative to PKS~2155-304 in equatorial coordinates 
(units are arc minutes).  Negative numbers indicate positions south and 
west of PKS~2155-304.

$^c$ Maximum brightness temperatures and 1$\sigma$ errors for Gaussian line 
shape fits with the central velocities and widths listed.  Upper limits 
(4$\sigma$) on T$_{max}$ between -400 and -100\kms\ are equal 
to 4 times the measured RMS baseline fluctuations when no line is present.  
The errors associated with detections are generally smaller than those in 
non-detection cases since the height of the Gaussian fit to the data is 
constrained by $\approx$20 independent data points.

$^d$ \ion{H}{1} column densities and errors (1$\sigma$) derived from the 
NRAO 21\,cm emission line measurements.  Upper limits are 4$\sigma$ 
estimates assuming FWHM = 30\kms.

$^e$ Broad, weak emission line may extend out to -130\kms.

$^f$ Weak (T$_{max} <$ 4$\sigma$) emission line for which the listed 
FWHM velocity width has additional uncertainties due to baseline removal.  
The limit on N(\ion{H}{1}) is the greater of the 4$\sigma$ detection limit 
(assuming FWHM = 30\kms) or the measured column density. 

\clearpage
\begin{deluxetable}{cccccccc}
\tablewidth{0pt}
\tablecaption{Equivalent Widths and Column Densities for the Mrk~509 
HVCs\tablenotemark{a}}
\tablehead{\colhead{Ion} & \colhead{$\lambda$(\AA)\tablenotemark{b}} 
	& \colhead{f\tablenotemark{b}}
	& \multicolumn{2}{c}{[\vlsr \ = $-$340 to $-$250\kms]} 
	& \multicolumn{2}{c}{[\vlsr \ = $-$250 to $-$170\kms]} 
	& \colhead{Source} \\
	& & & \colhead{W$_\lambda$(m\AA)} & \colhead{log\,N$_a$}
	& \colhead{W$_\lambda$(m\AA)} & \colhead{log\,N$_a$} & }
\startdata
\ion{C}{2}  & 1334.532  & $0.1278$	 & 70$\pm$17 &	13.63$\pm^{0.10}_{0.13}$   & 22$\pm$18	& $<$13.49  & 1 \nl
\nl
\ion{C}{4}  & 1550.770	& 0.0952	& 186$\pm$17 &	14.20$\pm$0.05	& 70$\pm$18  & 13.68$\pm$0.10 &	2 \nl
	    & 1548.195	& 0.1908	& 218$\pm$17 &	13.96$\pm$0.04	& 89$\pm$17  & 13.48$\pm$0.09 &	2 \nl
\nl
\ion{N}{1}  & 1199.550	& 0.1328	& 17$\pm$21  &	$<$13.58        & -20$\pm$21 &	$<$13.39      &	1 \nl
\nl
\ion{N}{5}  & 1238.821	& 0.1570	& 10$\pm$11  &	$<$13.18 	& 3$\pm$9    &	$<$12.98      &	2 \nl 
\nl
\ion{Si}{2} & 1304.370	& 0.0934	& 5$\pm$15   &	$<$13.40	& -6$\pm$13     & $<$13.27      &	1 \nl
	    & 1526.707	& 0.1155	& -5$\pm$23  &	$<$13.29	& 11$\pm$20	& $<$13.33	& 2 \nl
	    & 1190.416	& 0.2502	& 9$\pm$23   &	$<$13.29	& 20$\pm$21	& $<$13.38	& 1 \nl
	    & 1193.290	& 0.4991	& 10$\pm$23  & 	$<$13.00	& -2$\pm$21	& $<$12.84	& 1 \nl
	    & 1260.422	& 1.1150	& $<$35:\tablenotemark{d}    &	$<$12.65:$^d$	& \nodata	& \nodata	& 1 \nl
\nl
\ion{Si}{3} & 1206.500	& 1.6690	& 192$\pm$16 &	13.19$\pm$0.05	& 41$\pm$20	& 12.40$\pm^{0.16}_{0.25}$ &	1 \nl
\nl
\ion{Si}{4} & 1402.770	& 0.2553	& 93$\pm$18  &	13.44$\pm^{0.08}_{0.09}$ 	& 17$\pm$19	& $<$13.15 &	1 \nl
	    & 1393.755	& 0.5140	& 121$\pm$15 &	13.26$\pm$0.06	& 36$\pm$15	& $<$12.93	& 1 \nl
\nl
\ion{S}{2}  & 1250.584	& 0.0054	& 2$\pm$6    &	$<$ 14.27	& 0$\pm$5	& $<$14.12	& 2 \nl
	    & 1253.811	& 0.0109	& 3$\pm$5    &	$<$13.93	& 0$\pm$5	& $<$13.82	& 2 \nl
\nl
\ion{S}{3}  & 1190.208	& 0.0222	& 10$\pm$23  &	$<$14.36	& 23$\pm$20	& $<$14.41	& 1 \nl

\tablenotetext{a}{Equivalent widths and apparent column densities calculated 
using Eq. (1) in the text.  Uncertainties are 1$\sigma$ measurement errors.  
Limits are 2$\sigma$ estimates.}

\tablenotetext{b}{Wavelengths and f-values from the data compilation 
prepared by Morton (1991) with updates for the \ion{Si}{2} lines as listed by 
Savage \& Sembach (1996).}

\tablenotetext{c}{Source: (1) This paper; (2) Paper~I}

\tablenotetext{d}{Continuum placement for the \ion{Si}{2} $\lambda$1260 line 
at these velocities is subjective and depends strongly upon the properties 
of the nearby \ion{S}{2} $\lambda$1259 absorption and the derivative of the 
slope near the peak of the Ly$\alpha$ emission profile of the Seyfert galaxy.  
In our analysis we adopt the more conservative column 
density limit derived from the other \ion{Si}{2} lines at these velocities.}
\enddata
\end{deluxetable}

\clearpage
\begin{deluxetable}{cccccccc}
\tablewidth{0pt}
\tablecaption{Equivalent Widths and Column Densities for the PKS~2155-304 
HVCs\tablenotemark{a}}
\tablehead{\colhead{Ion} & \colhead{$\lambda$(\AA)\tablenotemark{b}} 
	& \colhead{f\tablenotemark{b}}
	& \multicolumn{2}{c}{[\vlsr \ = -330 to -190\kms]} 
	& \multicolumn{2}{c}{[\vlsr \ = -190 to -90\kms]} 
	& \colhead{Source} \\
	& & & \colhead{W$_\lambda$(m\AA)} & \colhead{log\,N$_a$}
	& \colhead{W$_\lambda$(m\AA)} & \colhead{log\,N$_a$} & }
\startdata
\ion{C}{4}  & 1550.770	& 0.0952 &  60$\pm$17	& 13.52$\pm^{0.10}_{0.15}$ &	54$\pm$14 & 13.48$\pm^{0.10}_{0.13}$ & 1 \nl
	    & 1548.195	& 0.1908 & 104$\pm$17	& 13.49$\pm^{0.06}_{0.08}$ &	114$\pm$13 & 13.52$\pm$0.05 &	1 \nl
\nl
\ion{N}{5}  & 1242.804	& 0.0782 & -1$\pm$13	& $<$13.40	& 6$\pm$11	& $<$13.42 &	1 \nl
	    & 1238.821	& 0.1570 & -1$\pm$12	& $<$13.05	& $<$29	& $<$13.14 & 	1 \nl
\nl	
\ion{Si}{2} & 1526.707	& 0.1155 & -10$\pm$18	& $<$13.18	& -5$\pm$14 & $<$13.07	& 1 \nl
	    & 1260.422	& 1.1150 & ($<$50)	& ($<$12.86)	& 72$\pm$18 & 12.73$\pm^{0.10}_{0.13}$  &	2 \nl
\nl
\ion{S}{2}  & 1250.584	& 0.0054 & 30$\pm$12	& $<$14.78	& 2$\pm$10 &	$<$14.46	& 1 \nl
	    & 1253.811	& 0.0109 &  2$\pm$12	& $<$14.20	& -10$\pm$10 &	$<$14.12	& 1 \nl
	    & 1259.519	& 0.0162 & 26$\pm$21	& $<$14.52	& -14$\pm$18 &	$<$14.19	& 2 \nl

\tablenotetext{a}{Equivalent widths and apparent column densities calculated 
using Eq. (1) in the text.  Uncertainties are 1$\sigma$ measurement errors.  
Limits are 2$\sigma$ estimates. Values in parentheses are estimates based 
on comparison of multiple line strengths as described in text (see also 
note c in Table~5)}

\tablenotetext{b}{Wavelengths and f-values from the data compilation 
prepared by Morton (1991) with updates for the \ion{Si}{2} lines as listed by 
Savage \& Sembach (1996).}

\tablenotetext{c}{Source: (1) This paper; (2) This paper, with reduced data 
courtesy of J. Stocke.}

\enddata
\end{deluxetable}

\clearpage
\begin{deluxetable}{ccccc}
\tablewidth{0pt}
\tablecaption{Summary of Adopted HVC Column Densities\tablenotemark{a}}
\tablehead{\colhead{Ion} & \multicolumn{2}{c}{Mrk~509}
	& \multicolumn{2}{c}{PKS~2155-304} \\
	& \colhead{(-283\kms)} & \colhead{(-228\kms)}
	& \colhead{(-256\kms)} & \colhead{(-140\kms)}}
\startdata
\ion{H}{1} (21\,cm) &	$<$17.69  & $<$17.69	& $<$17.73   & $<$17.73 \nl
\nl
\ion{C}{2}  & 13.63$\pm^{0.18}_{0.32}$ 	& $<$13.49   & \nodata & \nodata \nl
\nl
\ion{C}{4}  & $>$14.20$^b$	 & $>$13.68$^b$	& 13.50$\pm^{0.06}_{0.07}$ & 13.50$\pm^{0.06}_{0.07}$ \nl
\nl
\ion{N}{1}	    & $<$13.58	& $<$13.39	& \nodata	& \nodata \nl
\nl
\ion{N}{5}	    & $<$13.18	& $<$12.98	& $<$13.05	& $<$13.14 \nl
\nl
\ion{Si}{2}	    & $<$13.00	& $<$12.84	& $<$12.86$^c$ & 12.73$\pm^{0.10}_{0.13}$ \nl
\nl
\ion{Si}{3}	    & $>$13.19	& 12.40$\pm^{0.28}_{0.92}$ & \nodata	& \nodata \nl
\nl
\ion{Si}{4}	    & $>$13.44$^b$	& $<$12.93	& \nodata	& \nodata \nl	
\nl 
\ion{S}{2}	    & $<$13.93	& $<$13.82	& $<$14.20	& $<$14.12 \nl
\nl
\ion{S}{3}	    & $<$14.36	& $<$14.41	& \nodata	& \nodata \nl

\tablenotetext{a}{Adopted ultraviolet absorption line column densities 
derived from the values listed in Tables~3 and 4.  Errors have been converted 
to 2$\sigma$ estimates.  Limits are also 2$\sigma$ estimates.  
The H I column density limits are 4$\sigma$ estimates based on the residual 
RMS intensity in the baselines after removal of side lobe contamination.}

\tablenotetext{b}{This lower limit is based on the apparent column densities 
listed in Table~3.  It is consistent with the limits obtained by applying a 
single-component Doppler-broadened curve of growth to the measured equivalent 
widths.  The best fit value obtained from a curve of growth analysis is 
approximately 0.2 dex higher than the listed limit.}

\tablenotetext{c}{This limit on N(\ion{Si}{2}) is based upon the amount of
 high velocity \ion{Si}{2} that could be present within the core of the 
\ion{S}{2} $\lambda$1259.5 line given the relative equivalent widths and 
errors of the three \ion{S}{2} lines observed.  The value listed is for 
a single component with Doppler width b(\ion{Si}{2}) $>$ 5\kms. If 
b(\ion{Si}{2}) $<$ 5\kms, N(\ion{Si}{2}) could be 
larger than the listed value.  If b(\ion{Si}{2}) $\approx$ b(\ion{C}{4}) $\approx$
33\kms, then log\,N(\ion{Si}{2}) $\le$ 12.53.}

\enddata
\end{deluxetable}

\clearpage
\begin{deluxetable}{lcccccc}
\tablewidth{0pt}
\tablecaption{Properties of the \ion{C}{4}-HVCs Determined from 
Photoionization Model Results\tablenotemark{a}}
\tablehead{& \multicolumn{3}{c}{\ion{C}{4}-HVC @ -283\kms} 
	   & \multicolumn{3}{c}{\ion{C}{4}-HVC @ -228\kms} \\
	   & \colhead{[Z/H]\,=\,-1.0} & \colhead{[Z/H]\,=\,-0.5} & \colhead{[Z/H]\,=\,0.0}
	   & \colhead{[Z/H]\,=\,-1.0} & \colhead{[Z/H]\,=\,-0.5} & \colhead{[Z/H]\,=\,0.0}}
\startdata
log\,N(H) (cm$^{-2}$)	& 19.61  & 19.05	& 18.43	& 18.66 & 18.16	& 17.61 \nl
log\,N(H I) (cm$^{-2}$)	& 16.80  & 16.30	& 15.70	& 15.20 & 14.70	& 14.30 \nl
log\,$\Gamma$	        & -2.50  & -2.55	& -2.40	& -1.90 & -1.90	& -1.90 \nl
T (K)	                & 19,000 & 14,800 & 	9,400	& 20,000 & 19,200  & 12,300 \nl
D (kpc)	                & 108    & 28	 & 8.5	& 62 & 16	& 4.0 \nl
log\,n$_H$ (cm$^{-3}$)	& -3.87 & -3.82	 & -3.97	& -4.48 & -4.48	& -4.48 \nl
P/k (cm$^{-3}$ K)$^b$	& 5.9 & 5.1	 & 2.3	& 1.5 & 1.5	& 0.94 \nl
n$_H$/n$_{HI}$		& 640 & 560	 & 540	& 2900 & 2900	& 2000 \nl

\tablenotetext{a}{Sample CLOUDY photoionization model results for the 
\ion{C}{4}-HVCs toward Mrk~509.  Two examples that satisfy the available 
ionic information are provided for each cloud.  The input QSO-like ionizing 
spectrum has a Lyman limit mean intensity of J$_\nu$(LL) 
= 1$\times$10$^{-23}$ erg~cm$^{-2}$~s$^{-1}$~Hz$^{-1}$~sr$^{-1}$ with 
negligible intervening opacity, and the clouds are assumed to be plane 
parallel slabs irradiated on both sides.  A solar abundance pattern was 
adopted for the calculations with an overall metallicity given by [Z/H].  
For models where the metallicity is super-solar and some of the Si is 
incorporated into dust grains, the values of n$_H$, n$_H$/n$_{HI}$, and 
D drop by factors of 5--10 and the values of P/k increase by factors of 
2--5 (see \S5 for details).}

\tablenotetext{b}{P/k is take to be 2.3n$_H$T.  The factor of 2.3 is
appropriate for a fully ionized gas containing 10\% He by number.}

\enddata
\end{deluxetable}

\clearpage
\begin{deluxetable}{lcccccc}
\tablewidth{0pt}
\tablecaption{A Comparison of Observed Ionic Ratios to 
Theory\tablenotemark{a}}
\tablehead{\colhead{Ratio} & \multicolumn{2}{c}{Mrk~509 HVCs} 
        & \colhead{$<$Disk--Halo$>$\tablenotemark{b}}
        & \colhead{Hot Cooling Gas\tablenotemark{c}}
        & \multicolumn{2}{c}{Photoionized Gas\tablenotemark{d}} \\
	& \colhead{-283} & \colhead{-228} & &
	& \colhead{-283} & \colhead{-228}}
\startdata
\cutinhead{High/High Ion Ratios}
N(\ion{C}{4})/N(\ion{N}{5}) & $>$10.5  & $>$5.0  & 4.0$\pm$2.4  & 2.2--6.8 & 18--24 & 3.9--4.9 \nl 
N(\ion{C}{4})/N(\ion{Si}{4})& ($\sim$7.2)  & $>$5.6 & 3.8$\pm$1.9 & 3.0--4.7 & 10--11 & 49--85 \nl
N(\ion{Si}{4})/N(\ion{Si}{3}) & \nodata & $<$3.4 & $<$1 & $\sim$0.10 & 0.6--1.0 & 1.7--2.8 \nl
\cutinhead{High/Low Ion Ratios}
N(\ion{C}{4})/N(\ion{C}{2})	& $>$3.7	& $>$1.5 & $\ll$1	& $\sim$0.01	& 5.2--7.1	& $\sim$100 \nl
N(\ion{C}{4})/N(\ion{Si}{2})	& $>$10	& $>$3.5	& $\ll$1	& $\sim$0.10	& 77--88	& $>$1000 \nl
N(\ion{Si}{4})/N(\ion{Si}{2})	& $>$1.7	& \nodata	& $\ll$1	& $\sim$0.02	& 7.2--8.6	& 38--90 \nl

\tablenotetext{a}{This table provides a comparison of the observed ionic 
ratios for the Mrk~509 \ion{C}{4}-HVCs at -283 and -228\kms\ with the ratios 
typically observed in disk and halo gases and those predicted 
by theory.  Values in parentheses are based upon rough approximations 
obtained using a single component curve of growth to estimate the column 
densities.}

\tablenotetext{b}{High/High ion ratios are average results from Sembach \et
(1997) and Sembach \& Savage (1992).  High/Low ion ratios are estimates 
made for the low velocity gas toward Mrk~509.}

\tablenotetext{c}{These cooling Galactic fountain results are from 
Benjamin \& Shapiro (1998).  The model is for radiatively cooling gas with 
an initial temperature of 10$^6$ K.  The High/Low ion 
ratios depend upon the assumed stopping point for the cooling calculations; 
higher values would be obtained if the cooling flow is observed early in its 
evolution when much of the flow is still very hot.}

\tablenotetext{d}{Results for the photoionized gas models in Table~6 are 
listed.  These photoionized gas results depend upon the parameters of the 
model as described in the text.}
\enddata
\end{deluxetable}

\clearpage
\begin{deluxetable}{lcccccccc}
\tablewidth{0pt}
\tablecaption{An Inventory of HVC Detections and Non-Detections Along
Extragalactic Sight Lines for which \ion{C}{4} and Low Ionization
Absorption Line Observations are Available\tablenotemark{a}}
\tablehead{\colhead{Object} & \colhead{$l$} & \colhead{$b$}
	& \colhead{HVC} & \colhead{\vlsr} & \colhead{Ionized Gas}
	& \colhead{Neutral Gas} & \colhead{\ion{H}{1}}
	& \colhead{Ref.\tablenotemark{b}} \\
        & \colhead{(\degr)} & \colhead{(\degr)}
	& \colhead{Identification} & \colhead{(\kms)}
	& \colhead{HVC?} & \colhead{HVC?} & \colhead{HVC?}}
\startdata
PKS~2155-304	& \phn17.7	& $-$52.2	& GCN?	& $-$140	& \ion{C}{4}	& \ion{Si}{2}	& No$^c$	& 1 \nl
&&			& GCN?	& $-$256	& \ion{C}{4}	& No	& No$^c$	& 1 \nl
 \nl
NGC~5548	& \phn32.0	& +70.5	& \nodata	& \nodata	& No	& No	& No	& 3 \nl
 \nl
Mrk~509		& \phn36.0	& $-$29.9	& GCN?	& $-$228	& \ion{C}{4}, \ion{Si}{3}	& No	& No$^c$	& 1,2 \nl
&&			& GCN?	& $-$283	& \ion{C}{4}, \ion{Si}{3}, \ion{Si}{4}	& \ion{C}{2}	& No$^c$	& 1,2 \nl
 \nl
H~1821+643	& \phn94.0	& +27.4		& Outer Arm	& $-$120	& \ion{C}{4}	& \ion{Si}{2}, \ion{S}{2}, \ion{Mg}{2}	& Yes	& 4 \nl
&& 			        & Unknown	& $-$213	& \ion{C}{4}	& No	& No	& 4 \nl
 \nl
HD~5980 (SMC)	& 302.1	& $-$45.0	& SMC	& +150	& \ion{C}{4}, \ion{Si}{4}, \ion{N}{5}	&  Many ions	& Yes	& 11 \nl
 \nl
HD~36402(LMC)	& 277.8	& $-$33.0	& LMC	& +270	& \ion{C}{4}, \ion{Si}{4}	& Many ions	& Yes	& 12,13 \nl
&& 			        & Outer Galaxy?	& +130	& \ion{C}{4}, \ion{Si}{4} 	& Many ions 	& Yes	& 12,13 \nl
 \nl
SN~1987A(LMC)	& 229.7	& $-$31.9	& LMC	& +270	& \ion{C}{4}, \ion{Si}{4}	& Many ions	& Yes	& 14 \nl
&&   				& Outer Galaxy?	& +130	& \ion{C}{4}, \ion{Si}{4} 	& Many ions 	& Yes	& 14 \nl
 \nl
Mrk~205		& 125.5	& +41.7	& Complex C	& $-$147	& No	& \ion{Mg}{2}	& No	& 5 \nl
&& 				& 125+41-209	& $-$209	& No	& \ion{Mg}{2}	& Yes	& 5 \nl
 \nl			
NGC~3516	& 133.2	& +42.4 & \nodata	& \nodata	& No	& No$^d$	& No	& 3 \nl
 \nl
NGC~4151	& 155.1	& +75.1	& \nodata	& \nodata	& No$^e$	& No	& No	& 3 \nl
 \nl
NGC~3783	& 287.5	& +23.0	& 287+22+240	& +240	& No	& \ion{S}{2}, \ion{Si}{2}, Fe II	& Yes	& 6,7 \nl
 \nl
3C~273		& 290.0	& +64.4	& \nodata	& \nodata	& No	& No	& No	& 8 \nl
 \nl
Fairall~9	& 295.1	& $-$57.8	& Mag. Stream	& +170	& \ion{C}{4}$^f$	& \ion{Si}{2}	& Yes	& 3,9\nl
&& 				& Mag. Stream	& +210	& No	& \ion{Si}{2}	& Yes	& 3,9 \nl
 \nl
HD~156359$^g$	& 328.7	& $-$14.5	& Unknown	& +125	& \ion{N}{5}	& \ion{Si}{2}	& No$^h$	& 10 \nl

\enddata
\end{deluxetable}

\clearpage
\begin{center} TABLE~8 NOTES \end{center}

$^a$ This table does not include high velocity gas detections 
made by the GHRS in low resolution mode or the Faint Object Spectrograph 
(FOS) in high resolution mode.  For a summary of the FOS detections, see 
Table~8 of Savage \& Sembach (1996).   One distant Galactic star with 
detected high velocity gas is also included (HD~156359).

$^b$ References: 1) This paper; 2) Sembach \et 1995b - Paper~I; 
3) Savage \et 1997; 4) Savage \et 1995;  5) Bowen, Blades, \& Pettini 1995;  
6) Lu \et 1994a; 7) Lu \et 1998; 8) Savage \et 1993b; 9) Lu \et 1994b; 
10) Sembach, Savage, \& Lu 1995a; 11) Fitzpatrick \& Savage (1983); 
12) Savage \& deBoer (1981); 13) Chu \et (1994) ; 14) Savage \et 1989.

$^c$ For both the Mrk~509 and PKS~2155-304 sight lines, no high 
velocity \ion{H}{1} emission is observed directly along the sight line.  
However, \hhvcs\ exist within 2\degr\ of each sight line.  See text for a 
discussion of the relationship of these clouds with the observed 
HVC absorption features.

$^d$ The non-detection of low ionization, high velocity 
absorption toward NGC~3516 is based upon measurements of the weak 
\ion{P}{2} line near 1532\AA\ and the \ion{C}{1} lines near 1560\AA.  
The \ion{C}{1} lines are contaminated by redshifted \ion{C}{4} absorption 
associated with the Seyfert galaxy.  Better estimates 
of the low ionization, high velocity gas will require observations of 
stronger low ionization lines.

$^e$ The relatively small redshift of NGC~4151 ($z$ = 0.00332) 
results in a substantial overlap between the Milky Way and extragalactic 
\ion{C}{4} components seen along the path to this AGN.  
Here we assume the \ion{C}{4} components seen at \vlsr\ = -582, +146\kms,
and at higher velocity (Weymann \et 1997) are produced in the environment of 
NGC~4151 rather than in the Milky Way.

$^f$ According to Lu \et (1994b) both members of the 
\ion{C}{4} doublet reveal 2$\sigma$ absorption near 
+170\kms, the velocity of an \ion{H}{1}-HVC
toward Fairall~9.  In a mean profile 
formed by averaging the two \ion{C}{4} profiles, the 
\ion{C}{4} equivalent width between 100 and 200\kms\ 
is 95$\pm$35 m\AA, corresponding to a column density of log 
N(\ion{C}{4}) $\sim$ 13.5.

$^g$ HD~156359 is a distant ($\sim$11 kpc) halo star that lies 
approximately 2.8 kpc from the Galactic plane.  It is listed here for 
comparison with the extragalactic sight lines.  High velocity absorption 
associated with the +125\kms\ cloud has also been detected in 
\ion{C}{2}, \ion{Mg}{2}, \ion{S}{2}, \ion{Fe}{2}, and \ion{Si}{3} with the 
IUE satellite by Sembach \et (1991).  Although this HVC is seen in 
\ion{N}{5} with the GHRS, it is not detected in \ion{C}{4} or \ion{Si}{4} 
with the IUE.

$^h$ The nearest known \ion{H}{1}-HVC is 6\degr\ away  
(\#364, \vlsr\ = +115\kms\ in Wakker \& van Woerden 1991).  However, the 
region was mapped coarsely on a 2\degr$\times$2\degr\ grid.